\documentclass{ieeeaccess}
\usepackage{cite}
\usepackage{amsmath,amssymb,amsfonts}
\usepackage{algorithmic}
\usepackage{graphicx}
\usepackage{textcomp}
\usepackage{balance}
\usepackage[caption=false]{subfig}

\usepackage{bm}
\makeatletter
\AtBeginDocument{\DeclareMathVersion{bold}
\SetSymbolFont{operators}{bold}{T1}{times}{b}{n}
\SetSymbolFont{NewLetters}{bold}{T1}{times}{b}{it}
\SetMathAlphabet{\mathrm}{bold}{T1}{times}{b}{n}
\SetMathAlphabet{\mathit}{bold}{T1}{times}{b}{it}
\SetMathAlphabet{\mathbf}{bold}{T1}{times}{b}{n}
\SetMathAlphabet{\mathtt}{bold}{OT1}{pcr}{b}{n}
\SetSymbolFont{symbols}{bold}{OMS}{cmsy}{b}{n}
\renewcommand\boldmath{\@nomath\boldmath\mathversion{bold}}}
\makeatother

\def\BibTeX{{\rm B\kern-.05em{\sc i\kern-.025em b}\kern-.08em
    T\kern-.1667em\lower.7ex\hbox{E}\kern-.125emX}}

\begin{document}
\history{Date of publication xxxx 00, 0000, date of current version xxxx 00, 0000.}
\doi{10.1109/ACCESS.2024.0429000}

\title{Overview of NR Enhancements for Extended Reality (XR) in 3GPP 5G-Advanced}

\author{\uppercase{Margarita Gapeyenko}\authorrefmark{1},
\uppercase{Stefano Paris}\authorrefmark{1}, \uppercase{Markus Isomaki}\authorrefmark{1}, \uppercase{Boyan Yanakiev}\authorrefmark{1}, 
\uppercase{Abolfazl Amiri}\authorrefmark{1}, \uppercase{Benoist Sébire}\authorrefmark{1}, \uppercase{Jorma Kaikkonen}\authorrefmark{1}, 
\uppercase{Chunli Wu}\authorrefmark{1}, and \uppercase{Klaus I. Pedersen}\authorrefmark{1,2}}

\address[1]{Nokia Standards}
\address[2]{Department of Electronic Systems, Aalborg University, Aalborg, Denmark}

\markboth
{M. Gapeyenko \headeretal: Overview of NR Enhancements for Extended Reality (XR) in 3GPP 5G-Advanced}
{M. Gapeyenko \headeretal: Overview of NR Enhancements for Extended Reality (XR) in 3GPP 5G-Advanced}

\corresp{Corresponding author: Margarita Gapeyenko (e-mail: margarita.gapeyenko@nokia.com).}

\begin{abstract}
Extended reality (XR) is unlocking numerous possibilities and continues attracting individuals and larger groups across different business sectors. With Virtual reality (VR), Augmented reality (AR), or Mixed reality (MR) it is possible to improve the way we access, deliver and exchange information in education, health care, entertainment, and many other aspects of our daily lives. However, to fully exploit the potential of XR, it is important to provide reliable, fast and secure wireless connectivity to the users of XR and that requires refining existing solutions and tailoring those to support XR services. This article presents a tutorial on 3GPP 5G-Advanced Release 18 XR activities, summarizing physical as well as higher layer enhancements introduced for New Radio considering the specifics of XR. In addition, we also describe enhancements across 5G system architecture that impacted radio access network. Furthermore, the paper provides system-level simulation results for several Release~18 enhancements to show their benefits in terms of XR capacity and power saving gains. Finally, it concludes with an overview of future work in Release~19 that continues developing features to support XR services. 
\end{abstract}


\titlepgskip=-21pt

\maketitle

\section{Introduction}
\label{sec:introduction}

\PARstart{A}{doption} of extended reality (XR) across various sectors of daily life is continuously growing every \textcolor{black}{year~\cite{Metaverse_new, bl1}}. Consumers and professionals are able to choose from a range of devices and form factors, e.g., augmented reality (AR) glasses or virtual/mixed reality (VR/MR) helmets already today. Such devices together with new applications are unlocking new forms of education (e.g., virtual labs and lectures), improving online work environment, bridging people across the globe no matter the real distance, expanding coverage of health care, and other new use cases~\cite{bl2, bl3, XR_overview, XR_book, XR_book2}.

Besides devices and applications, one vital block to fully enjoy the possibilities that XR provides is wireless connectivity. It is of special importance to provide seamless connectivity to ensure the best user experience. For this, the 3rd Generation Partnership Project (3GPP) is taking an active role to continuously improve support of XR over NR radio access technology in 5G and beyond. \textcolor{black}{First, 3GPP identified various possible XR applications, defined a new 3GPP traffic model tailored for XR services and delivered several evaluation methodologies for capacity, power saving, coverage, and mobility (in Release~17 as well as earlier releases).} As the next step, 3GPP conducted a study to analyze XR over NR with the help of system level simulations~\cite{bl4}. The study demonstrated which additional improvements to NR could increase XR performance. Therefore, accounting for specifics of XR (described in more details in~\cite{bl4} and briefly summarized in Section~\ref{sec:traffic}), Release~18 introduced numerous enhancements across radio access and core networks. \textcolor{black}{Future research directions for 5G-Advanced standards were also outlined in~\cite{5G_advanced}.} 

In addition to 3GPP activities, industry, academia and various other stakeholders are actively exploring XR from different angles. For example, in~\cite{bl5} the paper explored positioning and communications for XR devices with the help of reconfigurable intelligent surfaces (RIS). In~\cite{bl6}, authors studied an issue related to three-dimensional (3D) orientations and 3D locations of user devices and provided an analysis of carrier phase measurements-based estimation approach. Further in~\cite{bl7}, authors considered resource allocation for XR users based on reinforcement learning techniques.

In~\cite{bl8} authors provided an overview of several power saving techniques and modeled a selected list of those to show the power saving gains for XR. Additional improvements for XR were demonstrated in~\cite{bl9} with the help of XR application feedback for loop back mechanism that adjusts XR traffic to network conditions. In~\cite{bl10}, the article showed performance of XR in presence of enhanced mobile broadband (eMBB) users, where adding eMBB users decreased the number of supported XR users.

\textcolor{black}{Furthermore, in~\cite{XR_rel17_paper1} authors summarized Release~17 study conducted in Radio Access Network (RAN) group 1 (RAN1) as well as outlined potential enhancements for future releases. In~\cite{XR_rel18_paper1} authors provided a summary of enhancements introduced in Release~18 for XR focusing mostly on RAN part. Our paper complements the earlier XR paper in~\cite{XR_rel17_paper1} and recent magazine paper in~\cite{XR_rel18_paper1} by considering particular enhancements that were agreed and specified in Release~18 and providing a comprehensive, tutorial like description of each of those enhancements with simulation results showing benefits of various enhancements.}

\textcolor{black}{Summarizing, in this paper, we provide a detailed overview of Release~18 NR enhancements introduced in Service and System Aspect (SA) and RAN working groups.} 

\textcolor{black}{The rest of the paper is organized as follows:}
\begin{itemize}
\item \textcolor{black}{In Section~\ref{sec:traffic}, we summarize main XR traffic characteristics to better understand the motivation of enhancements introduced in Release~18.} 
\item \textcolor{black}{In Section~\ref{sec:rel18_enhanc}, we provide a comprehensive overview of Release~18 XR enhancements.} 
\item \textcolor{black}{In Section~\ref{sec:sim_results}, a few selected Release~18 enhancements are analyzed with the help of system level simulations to demonstrate benefits of introduced enhancements.} 
\item \textcolor{black}{Further, in Section~\ref{sec:rel19_enhanc}, we proceed with an outlook of work continued in Release~19 that aims to improve XR performance even further.}
\item \textcolor{black}{Finally, in Section~\ref{sec:conclusion}, we provide a conclusion and discuss potential future directions.}
\end{itemize}

The key contributions of this paper are:
\begin{itemize}
\item The paper delivers a holistic overview of XR enhancements related to RAN and parts of core impacting RAN introduced in 5G-Advanced Release~18.
\item The paper continues with a system level analysis of selected Release~18 XR enhancements to demonstrate capacity and power saving gains.
\end{itemize}

\section{XR Traffic Model}
\label{sec:traffic}

This section provides an overview of 3GPP statistical XR traffic model to better understand the motivation and opportunities of the enhancements introduced in Release~18 and described in Section~\ref{sec:rel18_enhanc}. 

3GPP defined a new statistical traffic model specifically tailored to XR characteristics taking into account real traces provided by SA4 working group during Release~17~\cite{bl11}. The detailed traffic model description can be found in~\cite{bl4, bl12}, while in this paper we provide a summary of main components of the model. Depending on XR application, different types of traffic in both downlink (DL) and uplink (UL) direction may exist. Typically, for system level analysis of XR performance two main types of traffic are assumed: (i) video traffic and (ii) pose/control traffic. It is also possible that audio might be transmitted together with video as a separate stream, but as mentioned in~\cite{bl11}, the payload of such stream is rather small and can be neglected when conducting a simulation campaign.

\begin{figure}[!t]
 \centering
 \includegraphics[width=0.95\columnwidth]{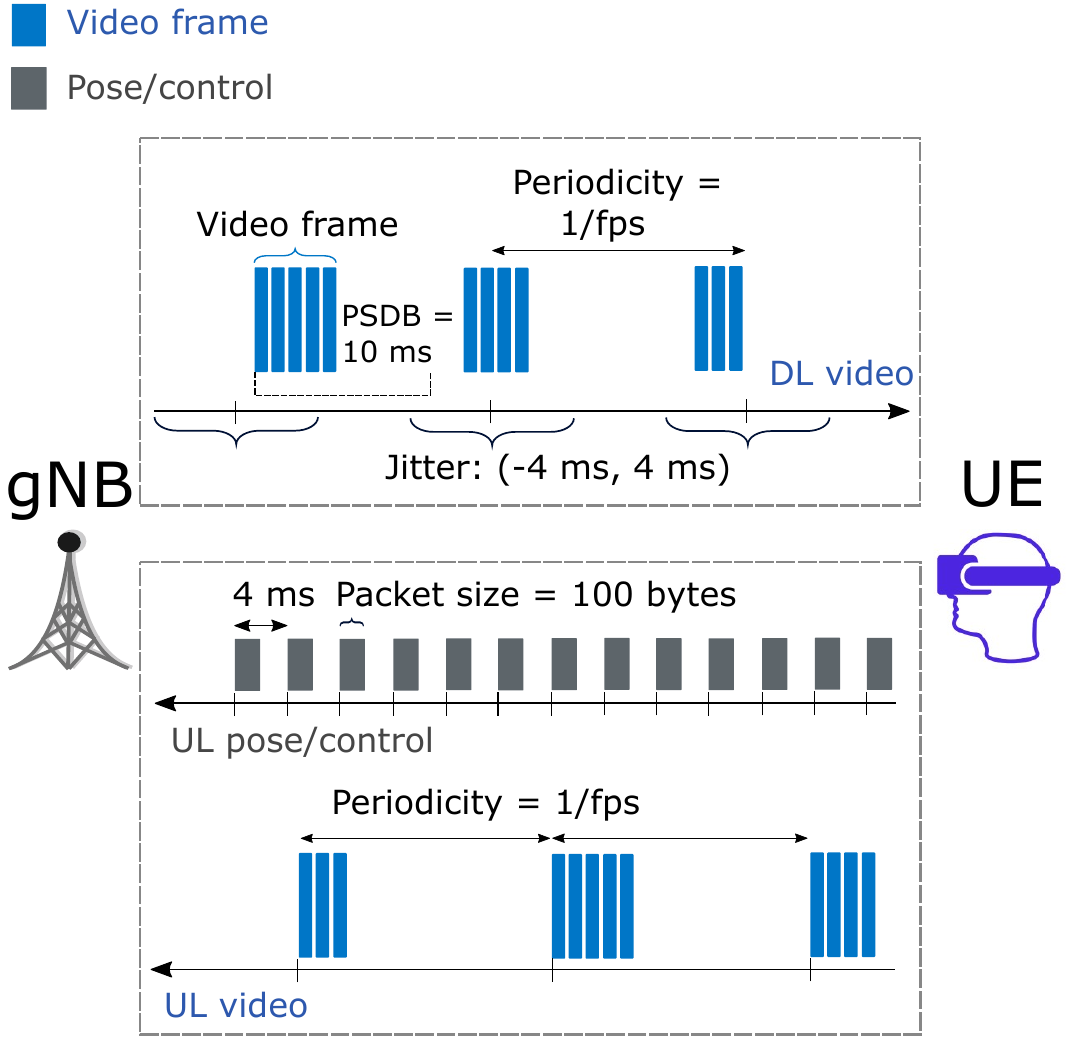}
 \caption{3GPP XR traffic model.}
 \label{fig:fig1}
\end{figure}

We start with a description of video traffic model. As shown in Fig.~\ref{fig:fig1}, video can be present in both directions, DL and UL, depending on the application that is running at user equipment (UE) side. The traffic model for video consists of three main components: video frame size, periodicity of frame arrivals, and jitter.  

According to Release~17 RAN1 working group study, the traffic model generates individual video frames with size following the truncated Gaussian distribution. The parameters for the video frames size generation can be found in Section~\ref{sec:sim_results} and~\cite{bl4, bl12}. According to the model, the average size of a video frame is around tens of kbytes.

The frames arrive in a sequence with periodicity equal to 1/[fps], where ``fps'' stands for ``frames per second''.  Therefore, e.g., for 60~fps, the frame rate assumed for the simulation purposes in Section~\ref{sec:sim_results}, time between frame arrivals is equal approximately to 16.67 ms, leading to non-integer periodicity between time arrival of the frames. 

\begin{figure*}[!t]
 \centering
 \includegraphics[width=0.95\textwidth]{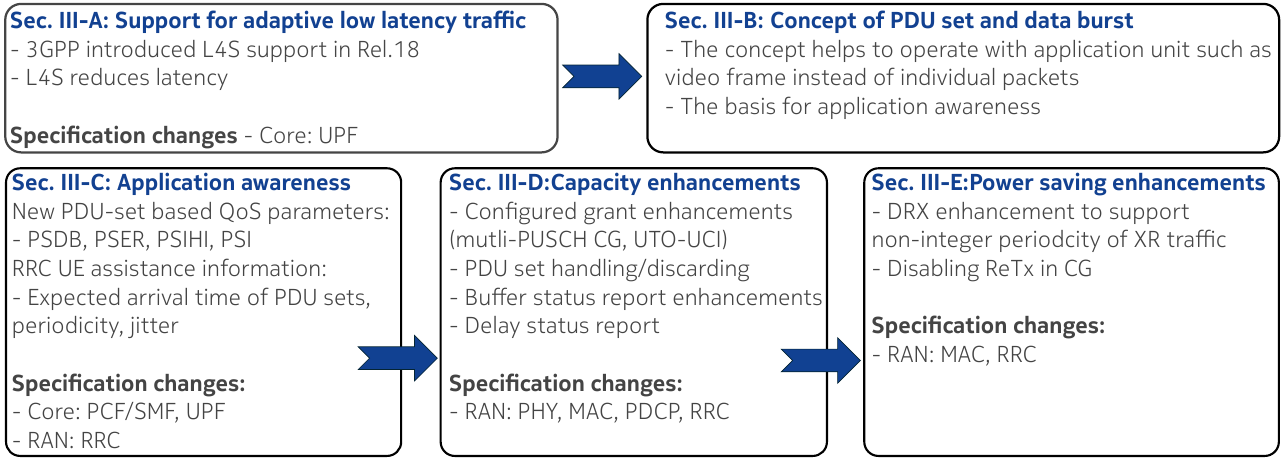}
 \caption{\textcolor{black}{Structure of Section~\ref{sec:rel18_enhanc} summarizing enhancements in Release 18 and highlighting specification impact.}}
 \label{fig:fignew}
\end{figure*}

The exact periodic frame arrival assumes there is no delay variation at any of the stages starting from video generation. However, in real applications, video is encoded with varying delay, and the delay induced by the network is not fixed either. All of the above leads to delay variation or ``jitter'' that affects the actual frame arrival at e.g., gNB. Therefore, the jitter is also considered in DL and modeled following the truncated Gaussian distribution with parameters given in Section~\ref{sec:sim_results}. 

When video is modeled in UL, for the sake of simplicity it is assumed that the jitter can be neglected, since video is generated closer to or at the UE and no transfer delay variation is introduced due to that.

The second type of traffic assumed for most of XR simulations in 3GPP is pose/control. The pose/control traffic model consists of packet size and packet arrival periodicity. The size of the packets is fixed and relatively small, e.g., 100~bytes as per 3GPP traffic model. The packets arrive periodically every few ms (4~ms as per 3GPP assumptions).

\section{XR Enhancements In Release 18}
\label{sec:rel18_enhanc}

In this section, we present an outlook of specified enhancements dividing them into five main groups: (i) \textcolor{black}{support for} adaptive low latency traffic; (ii) concept of PDU set and data burst; (iii) application awareness; (iv) capacity enhancements; and (v) power saving enhancements. \textcolor{black}{The introduced enhancements are aiming to reduce latency, increase number of simultaneously supported XR users as well as improving power saving performance of XR devices. We note that these enhancements were selected after RAN1 conducted a study in Release~17 and identified areas where further enhancements are needed to further improve XR performance.}

\textcolor{black}{For convenience, we provide Fig.~\ref{fig:fignew} to help with a navigation across enhancements described in this section where we also highlight particular protocols and layers in RAN and 5G core (5GC) impacted by introduced enhancements. Moreover, we include abbreviations used in the paper in Table~\ref{tab1} at the end of the paper}.


\subsection{Support For Adaptive Low Latency Traffic}

For interactive XR applications, latency is one of the most critical parameters affecting the user experience. In many cases users can tolerate temporary degradations of picture or audio quality as long as delay remains low. Such applications may support different quality levels with different bitrates, adapting to what is available in the network.
   
The network may aim to provide applications guaranteed quality of service (QoS) with a specific bitrate by admitting only as many users as it has capacity for while rejecting the rest. A more scalable approach is to dynamically divide the available capacity between the users as long as their applications can adapt to it. This can be done by traditional mechanisms, e.g., Transmission Control Protocol’s (TCP) classic congestion control reacting to packet loss. The problem with the traditional approach has been that the adaptation relies on packet loss and before losses the queuing latency already increases. This causes both loss and more disturbingly delay variation which the XR applications do not tolerate well especially when the delay increases albeit temporarily above their maximum delay requirements.

Low Latency Low Loss Scalable Throughput (L4S) Internet Service is a new Internet Engineering Task Force (IETF) standard attempting to provide consistent low network latency for applications adhering to a new congestion control mechanism~\cite{bl15, L4S_2}. The applications still need to adapt to the available bitrate, but their network delay stays low.

L4S is based on the following principles~\cite{L4S_Nokia}:
\begin{itemize}
\item In the network nodes: Isolation of L4S vs traditionally congestion-controlled traffic flows and providing early and granular congestion notification by marking the Explicit Congestion Notification (ECN) bits in IP header. Performing the marking early means that it is started with low probability already before serious congestion emerges. This way congestion related queuing delay or packet loss can be avoided for the L4S compliant traffic flows. 
\item	In the endpoints (IP hosts): The Receiver provides the sender feedback on ECN markings performed by the network and the Sender adapts its bitrate (congestion window) in L4S compliant way based on the feedback. How fast the Sender decreases its bitrate depends on the portion of the packets the network has marked. If the network has marked no packets, the Sender can gradually increase its bitrate until markings appear again. This way there is always an equilibrium between how much the Sender transmits and how much the network can deliver before queuing delays start to increase.
\item	In the applications: Adapting to the available bitrate by means of, e.g., altering the video quality or frame rate.
\end{itemize}

The feedback mechanism and congestion control are embedded in transport protocols such as QUIC, TCP or Real Time Transport Protocol (RTP) with RTP control protocol (RTCP) feedback. 

\begin{figure}[!t]
 \centering
 \includegraphics[width=0.95\columnwidth]{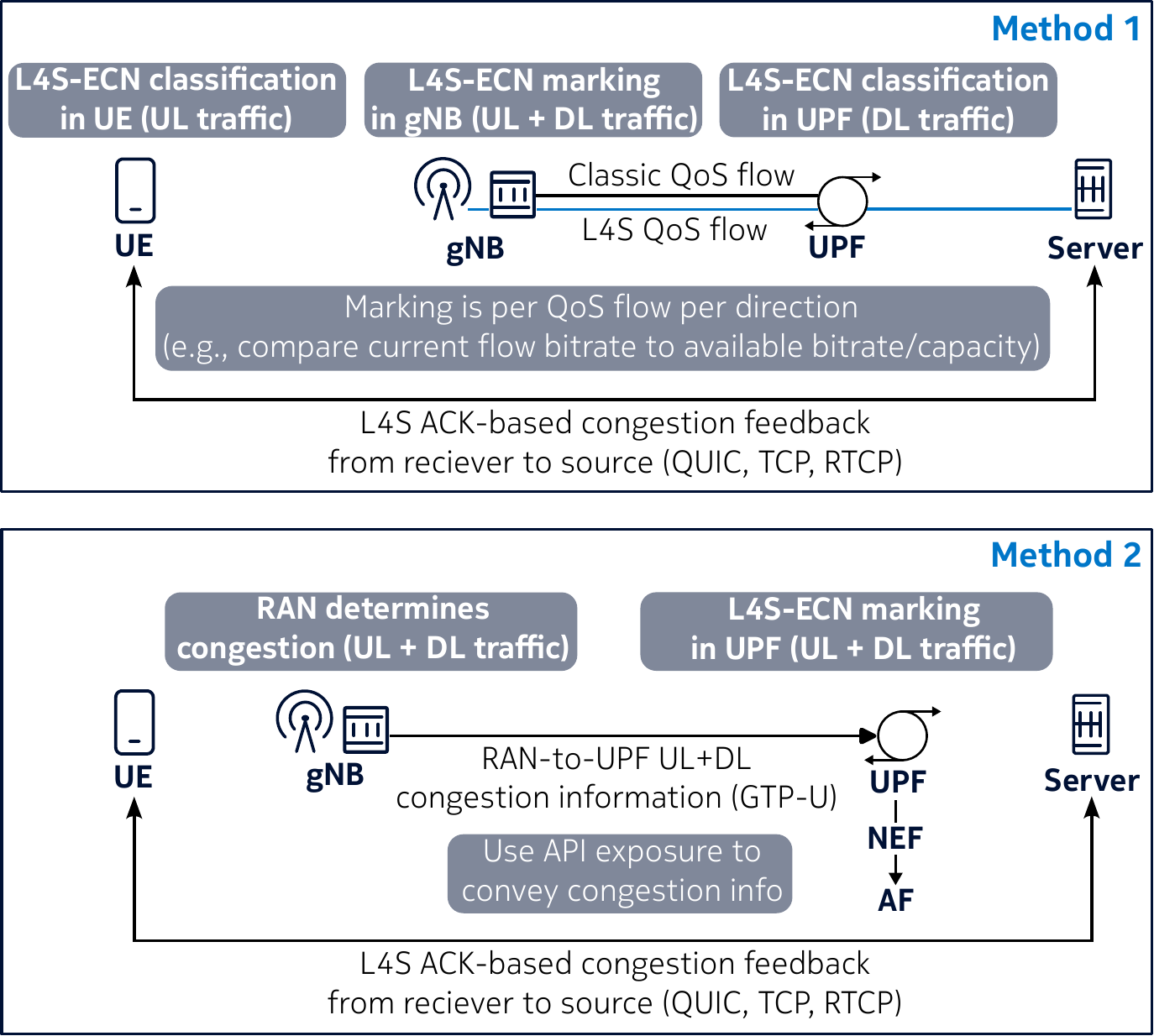}
 \caption{Illustration of L4S methods specified in 3GPP. \textcolor{black}{Method 1: RAN performs the ECN marking. Method 2: RAN provides the ECN marking probability to UPF, which performs the marking on RAN’s behalf.}}
 \label{fig:fig3}
\end{figure}

To improve support for adaptive low latency traffic in compliance with L4S introduction in IP networks in general, 3GPP introduced L4S support for 5G in Release 18~\cite{5GS_XR_Rel18}. The required isolation of L4S traffic from the ``classic'' traffic is achieved by separating the L4S traffic flows into one or more dedicated QoS flows. \textcolor{black}{The flow level separation allows the network to share capacity between L4S and classic traffic and also between users in a flexible manner depending on the QoS parameters of each QoS flow.} For the L4S QoS flows the 5G system (5GS) performs the ECN marking according to the requirements of the IETF standard. Two methods have been standardized:
\begin{enumerate}
\item RAN performs the ECN marking as illustrated in Fig.~\ref{fig:fig3} under Method 1.
\item	RAN determines the marking probability (percentage of packets to be marked) and provides that over GPRS Tunnelling Protocol for the user plane (GTP-U) to User Plane Function (UPF) which performs the marking as shown in Fig.~\ref{fig:fig3} under Method 2. 
\end{enumerate}

This is done for both DL and UL traffic. Exactly how the RAN performs the marking or determines the marking probability is left for RAN implementation but should reflect the bitrate (resources) available for the QoS flow compared to the current rate of the flow and should be done in such a way that the congestion for the QoS flow does not yet arise. 
\textcolor{black}{Performance of L4S is described in~\cite{L4S_Nokia}, where authors demonstrated simulation results showing significant improvements in queuing delay when applying L4S as compared to other mechanisms even in very congested scenarios.}

\subsection{Concept of PDU Set And Data Burst}

Media streams are transported over the network in application layer data units following a specific structure. For instance, a video stream consists of different types of frames (see Fig.~\ref{fig:fig1}) which may further comprise of slices or tiles. Each of these are transported over the network by fragmenting them into a set of Internet protocol (IP) packets. They may have a different importance wrt. quality of video stream based on their type or characteristics known to the application. Similarly other media streams, such as haptics, may consist of data units transported by a set of IP packets. Audio or pose information on the other hand typically consist of encoded samples that are small enough to fit into a single IP packet. 

To exploit this structuring, 3GPP has defined a concept of a PDU set as ``one or more PDUs carrying the payload of one unit of information generated at the application level''~\cite{bl14}. Another concept that 3GPP defined related to the traffic is Data Burst. Data Bursts is a set of multiple PDUs generated and sent by the application in a short period of time. Periodicity refers to the time interval between Data Bursts. 

\begin{figure}[!h]
 \centering
 \includegraphics[width=0.95\columnwidth]{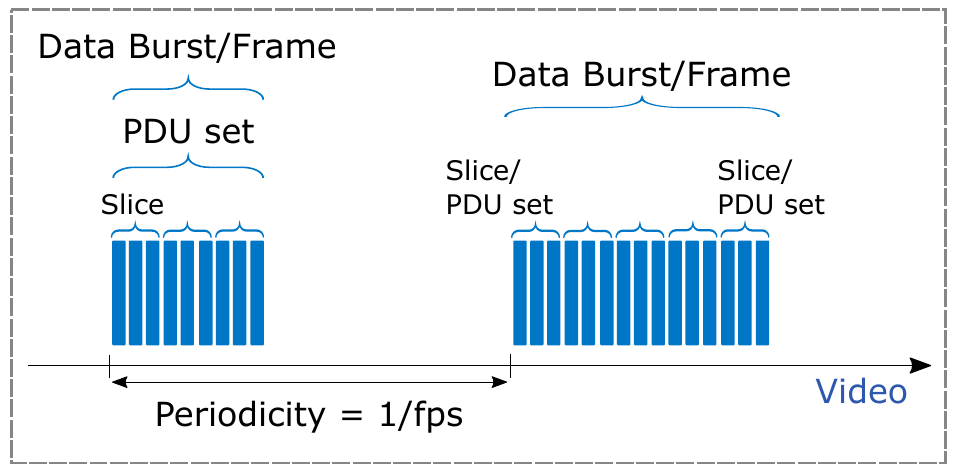}
 \caption{Relation between data burst, frames, and PDU sets.}
 \label{fig:fig4}
\end{figure}

The relationship between a Data Burst, Frame and a PDU Set depends on the application. In one example, the Data Burst in a video stream is a single frame which is also treated as a single PDU set (1 Frame = 1 Data Burst = 1 PDU set) as illustrated in Fig.~\ref{fig:fig4}. In another example, the video is encoded as slices with each frame consisting of multiple slices and the PDU sets are formed based on slices as shown in Fig.~\ref{fig:fig4}. In that case 1 Data Burst = 1 Frame =  multiple (e.g., 10) PDU sets. The number of slices per Data Burst or frame can vary per frame, and the number and size of IP packets per PDU set can vary per slice, so this information can only be provided along with the packets or PDUs themselves.

\begin{figure*}[!t]
 \centering
 \includegraphics[width=0.85\textwidth]{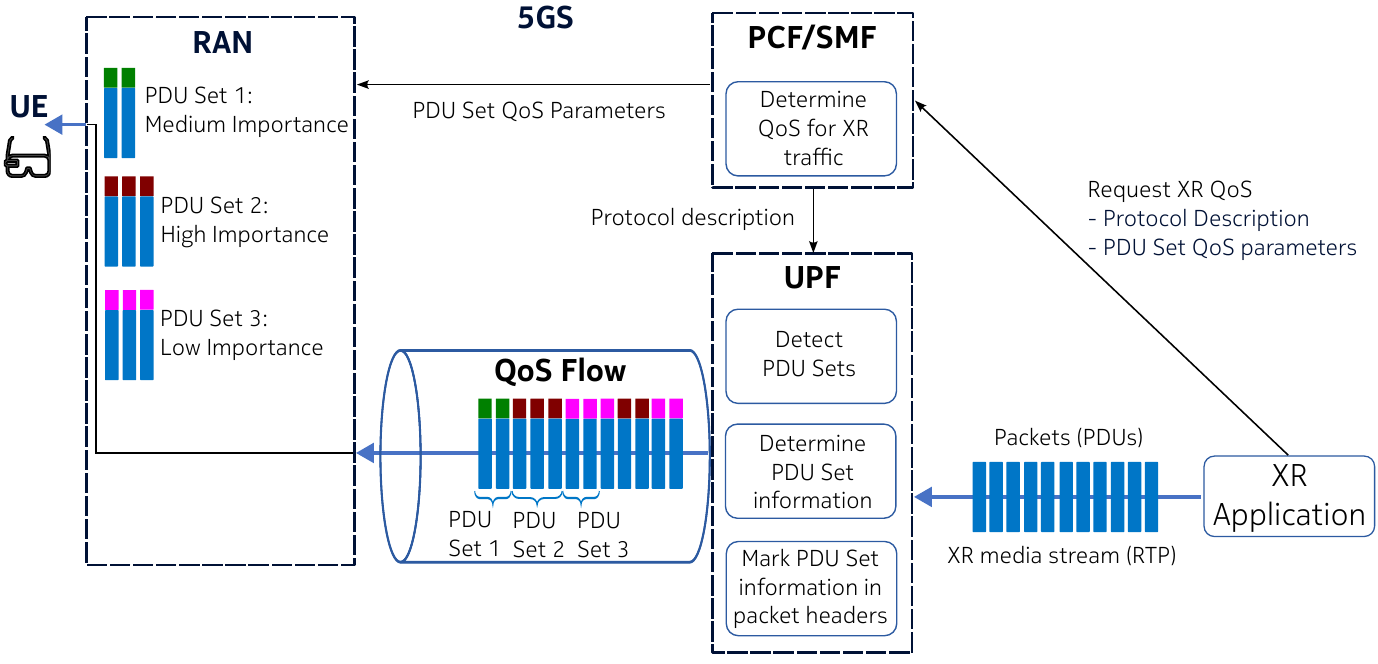}
 \caption{Example of application awareness: Information about PDU sets is transferred across 5GS.}
 \label{fig:fig5}
\end{figure*}

\subsection{Application Awareness}

Traditionally the 5G System treats User Plane traffic in terms of PDUs and traffic flows. PDUs are individual IP packets (or Ethernet frames but in XR use cases IP based traffic is assumed) while traffic flows comprise all IP packets exchanged with two endpoints identified by endpoint IP addresses, transport protocol and transport protocol port numbers. In case of XR, the application may transmit its media streams (video, audio, haptics) and other data (e.g., pose information) using one or multiple traffic flows. In traditional QoS handling all PDUs of a traffic flow are mapped to a specific QoS treatment based on a set of parameters that apply to each PDU alike.

One of the enhancements introduced in Release~18 to support XR was to make traffic handling application aware by treating the traffic according to the application layer structures and patterns in the RAN. This is explained in detail in the following:

\subsubsection{PDU Set Based QoS Handling}

The general aim of PDU set based QoS handling is to provide QoS on the level of data units meaningful on the application layer such as video frames or slices by treating them as PDU sets in the network. This means that instead of providing QoS with targets for individual PDUs such as Packet Delay Budget (PDB) or Packet Error Rate (PER), PDU sets are handled in an ``integrated'' manner by subjecting the entire PDU set to new PDU set specific QoS parameters, which are defined as follows~\cite{bl14, 5GS_XR_Rel18}: 
\begin{itemize}
\item PDU Set Delay Budget (PSDB) defines an upper bound for the delay that a PDU set may experience for the transfer between the UE and the N6 termination point at the UPF, i.e., the duration between the reception time of the first PDU (at the N6 termination point for DL or the UE for UL) and the time when all PDUs of a PDU set have been successfully received (at the UE for DL or N6 termination point for UL).
\item	PDU Set Error Rate (PSER) defines an upper bound for the rate of PDU sets that have been processed by the sender of a link layer protocol (e.g., RLC in RAN of a 3GPP access) but that are not successfully delivered by the corresponding receiver to the upper layer (e.g., PDCP in RAN of a 3GPP access). Thus, the PSER defines an upper bound for a rate of non-congestion related PDU set losses.
\item	PDU Set Integrated Handling Indicator (PSIHI) indicates whether all PDUs of the PDU set are needed for the usage of the PDU set by the application layer on the receiver side.
\end{itemize}

In addition, it is possible to handle PDU sets in ``differentiated'' manner based on PDU Set Importance (PSI), which identifies the relative importance of a PDU set compared to other PDU sets within a QoS Flow and can be used, e.g., for PDU set level packet discarding in presence of congestion. In other words, this information can be used to discard less important PDUs, e.g. PDUs with marginal impact on user experience, to alleviate congestion in the radio link. 

For QoS flows designated for PDU set based QoS handing, the Core Network will provide one or more of the PDU set QoS parameters to RAN, in which case PSDB will supersede PDB and PSER will supersede PER. The PDU set QoS parameters are direction specific meaning they may have separate values for UL and DL direction. The original source of these values is the application which provides them to the Core Network either via a request to Network Exposure Function or by manual configuration. In general, PDU set information can be used for the scheduling decisions at the gNB (described in more details in Section~\ref{sec:capacity_enh}) to consider the inter dependencies among PDUs within a set opposed to legacy implementations that only relied on per-packet information for scheduling.

For DL traffic the UPF is required to identify the PDU sets and their characteristics, most importantly the importance (as illustrated in Fig.~\ref{fig:fig5}) and size of the PDU set. The identification is based on transport or application protocol header information in the IP packets sent by the application. 3GPP SA4 has defined an RTP header extension for this purpose but other protocols can be supported too. The UPF can be provided with a Protocol Description instructing it how to do the identification for a particular traffic flow. Once it has done the identification the UPF includes the information with every PDU (IP packet) in a GTP-U header to the RAN. This way the RAN becomes aware of the PDU sets and their characteristics without needing to perform the application protocol specific identification by itself.  In Fig.~\ref{fig:fig5}, we show an example of how information about the PDU sets is transferred across 5GS.

PDU sets can similarly be identified for UL traffic in the UE. In this case the Protocol Description is provided to the UE. Similar to the PDU set QoS parameters, the source of the Protocol description for both DL and UL traffic is the application which can provide it to the Core Network.

In addition to identifying the PDU sets, the UPF for DL direction and UE for UL direction can detect the ending of a Data Burst, i.e., the last packet of the burst. The detection is based on transport or application layer header information in the packets, for instance the RTP header extension defined by 3GPP SA4. For DL packets, the UPF inserts the End of Data Burst (EoDB) indication in the GTP-U header of that packet. This way the RAN knows accurately when each data burst ends, even if the data burst size in number of bytes or packets can be highly variable and unpredictable.

\subsubsection{Mapping of PDU SET to QoS and DRB}

As explained, the UPF identifies the PDU sets and maps them to QoS flows for which PDU set based QoS handling is requested. The QoS flows are further mapped onto Data Radio Bearers (DRB) by the RAN for the integrated and differentiated handling over the radio interface. The label ``L:M:K'' is defined by 3GPP to indicate the mapping configurations, where L, M, and K indicate the number of PDU sets, QoS flows, and DRBs, respectively. Only a limited number of mapping configurations among PDU sets, QoS flows, and DRBs are specified by 3GPP since the information to detect PDU sets is not any more easily accessible or unavailable as information is passed down across the layers of the 5GS and RAN. In particular, the specified mapping configurations are: 1:1:1, N:1:1, and N:N:1. Alternative 1:1:1 provide the highest degree of control on the QoS treatment of PDU sets since a dedicated path from PDU set to radio resources is dedicated to the application-layer stream, whereas Alternative N:N:1 and N:1:1 simplify the coordination and synchronization of streams mapped onto multiple PDU sets  since PDUs in the same QoS Flow or DRB are treated with the same priority. Note however that N:1:1 enables end-to-end coordination within the whole 5GS whereas N:N:1 limits the coordination to the RAN portion of the 5GS.

\subsubsection{RRC UE Assistance Information}

Characteristics of UL data traffic can be hardly measured by the gNB since multiple flows may be multiplexed in the low layer radio protocols. Additionally, the gNB requires to get this assistance information as soon as it becomes available at the UE in order to use it for radio resource management (RRM) operations. While Time Sensitive Communication Assistance Information (TSCAI)~\cite{TSCAI} was extended in Release 18 to include XR traffic characteristics like periodicity and jitter~\cite{bl14}, using this framework for handling UL traffic is not a viable solution. This would require waiting until the SMF has coordinated with UPF, which inspects the XR traffic, to assist RAN, which is not a best solution for real time operations such as uplink scheduling. To solve this issue, 3GPP RAN2 has extended the RRC UE Assistance Information (UAI) to include important XR traffic characteristics such as the expected arrival time of PDU sets, periodicity, jitter, and PSI carried over a certain DRB and QoS flow. To minimize the reporting overhead in UL, the network can configure a prohibit timer that limits the pace at which the UE can provide such information.

\textcolor{black}{Benefits from application awareness can be observed from Section~\ref{sec:sim_results} where knowledge about application unit can further improve capacity and increase power saving.}

\subsection{Capacity Enhancements}
\label{sec:capacity_enh}

Capacity is considered to be one of the important key performance indicators (KPI) for XR services. To increase the number of supported users while ensuring each XR user is satisfied in terms of packet delay budget and packet error rate, 3GPP developed a number of enhancements described in more details below.

\subsubsection{Configured Grant (CG) Enhancements}

\begin{figure}[!b]
 \centering
 \includegraphics[width=0.95\columnwidth]{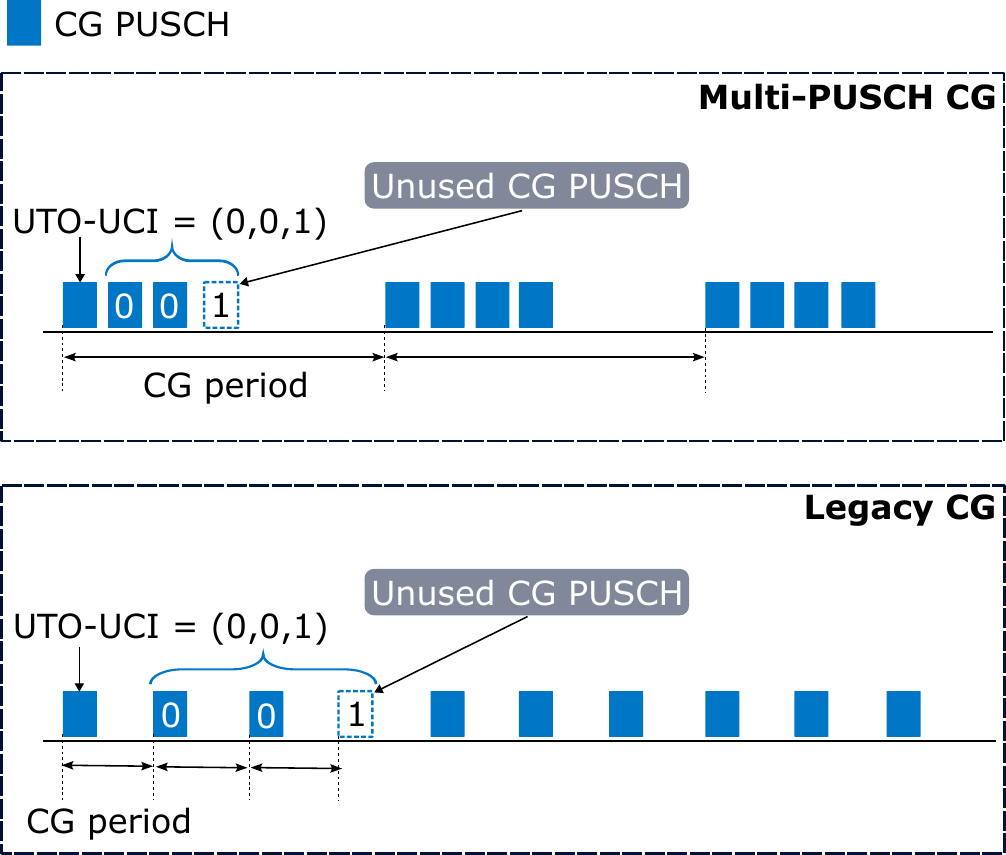}
 \caption{Release 18 XR enhancements introduced for configured grant: multi-PUSCH and UTO-UCI.}
 \label{fig:fig6}
\end{figure}

Some XR applications, e.g., augmented reality, require high data rate not only in DL but also in UL direction. According to studies conducted in~\cite{bl12}, number of XR users supported in UL can be quite limited if an application requires to convey a video traffic in UL direction. Thus, it is important to increase limited time to successfully transmit and receive the data. Increased scheduling delay may decrease the time left for the 
packet to be delivered successfully before the packet delay expires that will lead to decreased number of supported UEs. 

One possible solution to decrease the scheduling delay is configured grant (CG). Configured grant is a type of UL scheduling where gNB does not need to send a dedicated control signal for every transmission~\cite{5G_CG}. The parameters for scheduling are configured in RRC. There are two types of CG: (i) type 1 CG, where UE starts CG transmission as soon as RRC configuration/reconfiguration is processed at UE side; and (ii) type 2 CG, where CG configuration is activated/deactivated at UE side with a downlink control information (DCI).

Besides tight PSDB, video frames are also characterized by large and varying size where multiple transmission occasions are required to transmit one video frame (see Section~\ref{sec:traffic}). Therefore, in order to minimize time needed to transmit one video frame over multiple transmission occasions, the available transmission occasions shall be consecutive in time or with minimum gaps between those. These can be realized by two possible solutions: (i) CG periodicity is the smallest possible periodicity that allows transmitting one video frame over consecutive available transmission occasions or with minimum possible gap between transmission occasions (see Fig.~\ref{fig:fig6}, Legacy CG) or (ii) one CG periodicity shall consist of multiple consecutive transmission occasions (see Fig.~\ref{fig:fig6}, multi-PUSCH CG). The latter is already possible in unlicensed bands and Release 18 proposed the same solution to be applied in licensed bands with minimum modifications. The feature called multi-PUSCH (Physical uplink shared channel) CG allows to configure multiple consecutive transmission occasions during one CG periodicity as also shown in Fig.~\ref{fig:fig6}. 

In both cases, whether legacy CG with short periodicity or multi-PUSCH CG is employed, there is a problem that some resources might not being utilized. When one frame is transmitted over multiple transmission occasions, there will be a certain gap until the next frame arrival, thus all transmission occasions until the next frame arrival will not be needed in case of legacy CG. Multi-PUSCH CG can at some extent solve the problem, however, the exact frame size is unknown at the time of configuring the number of PUSCH per CG period, and it always varies from frame to frame, making it impossible to predict the exact number of resources needed. Thus, if the number of multi-PUSCH in one CG period is larger than required number of PUSCHs to transmit a video frame, some of the resources will be wasted. Therefore, the second feature called unused transmission occasion uplink control information (UTO-UCI) was specified in Release~18. 

UTO-UCI allows UE to indicate to gNB which of the upcoming occasions are going to be unused so gNB can re-use those for other users. The UTO-UCI is a bitmap that indicates for the next N consecutive transmission occasions whether those are unused (``1'') or not unused (``0'') as demonstrated in Fig.~\ref{fig:fig6}.

\textcolor{black}{Performance evaluation of CG enhancements can be found in Annex B.1.6 of TR 38.835~\cite{bl13}, where e.g., possibility to indicate that some of CG occasions as not used by UE, demonstrate capacity gains around 11\%-95\% as compared to dynamic grant scheduling.}

\begin{figure}[!b]
 \centering
 \includegraphics[width=0.95\columnwidth]{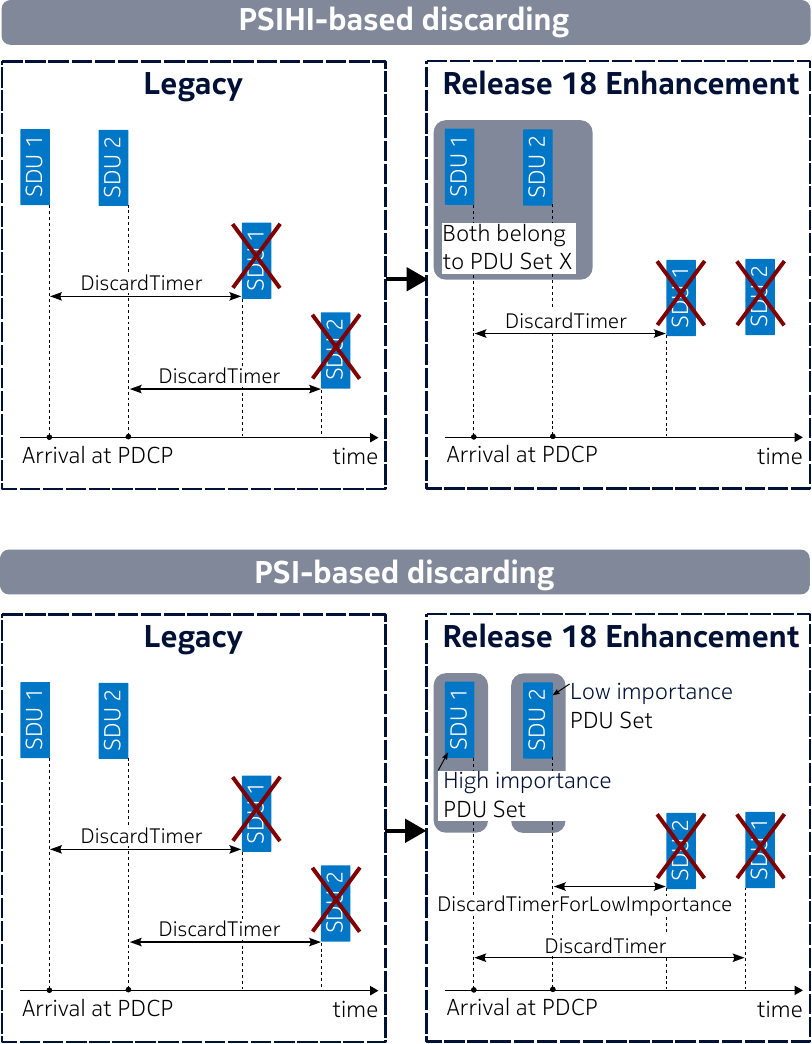}
 \caption{Illustration of discarding based on PDU set integrated handling indicator (PSIHI) and PDU set importance (PSI).}
 \label{fig:fig7}
\end{figure}

\subsubsection{PDU Set Handling/Discarding}

\begin{figure*}[!t]
 \centering
 \includegraphics[width=0.95\textwidth]{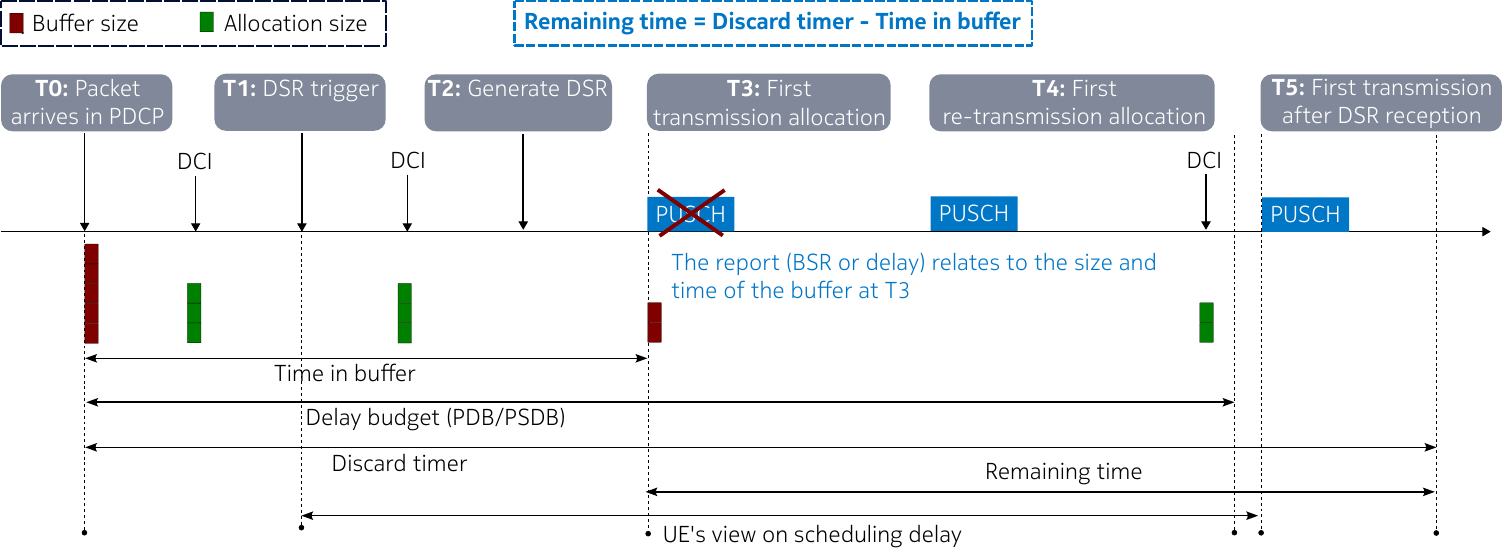}
 \caption{Illustration of DSR timing diagram.}
 \label{fig:fig8}
\end{figure*}

The notion of PDU set introduced above enables the RAN to identify the PDUs which carry content that the application processes as a single unit. This contributes to the optimization 
of radio resources with the help of discarding when combined with two other pieces of information: 
\begin{enumerate}
\item PSIHI-based discarding (illustrated in Fig.~\ref{fig:fig7}): with the PSIHI, the transmitter is able to tell that when one PDU of a PDU set is lost, the remaining PDUs of that PDU set can immediately be considered as no longer needed by the application and can thus be discarded to free up radio resources.
\item PSI-based discarding (illustrated in Fig.~\ref{fig:fig7}): with the PSI, the transmitter can decide to prioritize high priority PDU sets and discard low importance ones in case of congestion, also to free up radio resources. 
\end{enumerate}

In DL, the PSIHI is signalled to the gNB by the Session Management Function (SMF) as a PDU set QoS Parameter of the QoS flow profile, while the PSI is carried in the GTP-U headers. 

In UL, if the UE is configured with \textit{pdu-SetDiscard}, it discards all packets in a PDU set when one PDU belonging to the set is lost. The setting of \textit{pdu-SetDiscard} is done by the gNB based on the PSIHI received from the SMF. Regarding the PSI, and the identification of PDU set in general, this is left to UE implementation but when possible for a QoS flow, the UE informs the gNB via the UE Assistance Information.

\subsubsection{Buffer Status Report/Delay Status Report}

Buffer status report (BSR) is a key message UE sends to gNB informing of the availability and volume of data in UL. This is done by first calculating the total volume of data available in the UE buffer in bytes, per Logical Channel Group (LCG). Then the UE reports an index from a Buffer Status (BS) table known to both UE and gNB, where the index points to some number of bytes. When picking the reported index, the UE always rounds up to the nearest index which corresponds to an equal or higher number bytes to the actual byes in the buffer. This means that the UE almost always reports more bytes than it actually has.

To enhance the scheduling efficiency of UL resources, a finer granularity buffer status (BS) table for buffer status report has been introduced, with narrower range than legacy. It is mainly intended for XR traffic, but it is not restricted to XR. The network may configure each Logical Channel (LCH) to use the refined BS table and the UE will use it, if current buffer status falls into the range of the refined BS table. Otherwise, the UE falls back to the legacy table. BSR using the refined BS table is called as Refined Long BSR and it could be indicated via the Refined BSR MAC CE. The table used for each LCG can be indicated, so that the BS table can be dynamically selected by the UE based on whether the current buffer status falls into the range of the Refined BS table or not.

Additionally, to ensure timely scheduling and balancing the priority among the UEs, Delay Status Report (DSR) has been introduced. DSR allows the UE to report the amount of data is with short remaining delay budget and the associated remaining time in buffer before discard. The gNB can configure a threshold per LCG for the UE so that DSR is triggered and reported as long as there is data (that has not been reported before) with remaining time before discard becomes below the threshold. In the DSR, the UE will indicate both the smallest remaining time and the total buffered data with remaining time below the threshold for each LCG as also illustrated in Fig.~\ref{fig:fig8}. In Fig.~\ref{fig:fig8}, it is shown that a packet arrives at T0. At T1 the packet’s discard timer crosses the DSR trigger threshold – a gNB configured parameter. The two DCI messages shown on either side of T1 indicate that an UL grant can come before or after this trigger and there will be no change in operations. At T2 the DSR message is generated with remaining time value filled in the DSR MAC CE referring to the remaining time at T3 – the scheduled transmission slot/symbol. The PUSCH occasions indicate that upon PUSCH failure and retransmission via a grant in T4, the DSR timing and data volume received at T5 are uniquely identifiable by referring to the first transmission grant at T3. Note that remaining time is based on the discard timer configured by the gNB and not on the PSDB. Those can be the same, but they can also be different.

\textcolor{black}{Simulation results for enhanced BSR and DSR can be found in Section~\ref{sec:sim_results} and additionally from Annex B.1.8 TR 38.835~\cite{bl13}, where improving the granularity of BSR table improved radio resource distribution and that in turn increased capacity further.}

\subsection{Power Saving Enhancements}
\label{sec:ps_enh}

Another area of improvements that is also important for XR devices is power saving enhancements. A typical XR device form factor does not allow to have a large battery and the use cases requires the devices are not over heated while being used by a user. Thereby, power saving was identified as another important area to be enhanced.

To enable more power efficient UE operation in Connected mode, NR supports Discontinuous Reception (DRX) operation. DRX, when configured to the UE by network, provides UE opportunities to save power, by enabling UE to move the receiver to more power efficient state when data scheduling is not expected~\cite{bl16}. This is achieved by configuring UE with periodic time windows, during which UE is always expected to monitor Physical downlink control channel (PDCCH) and with an inactivity timer to determine the time for which PDCCH is monitored after receiving a PDCCH scheduling a new data transmission.
UE is configured with an inactivity timer, that is reset/started every time when a PDCCH scheduling a new data transmission is received and UE is required to monitor the PDCCH while the inactivity timer is running. Once the inactivity timer has expired, UE is required only periodically monitor PDCCH at least for a configured duration. The periodicity and the time location of the periodic PDCCH monitoring is set by configuring the starting sub-frame and slot and after the duration has expired, UE can stop PDCCH monitoring if it has not been scheduled.

\begin{figure}[!b]
 \centering
 \includegraphics[width=0.95\columnwidth]{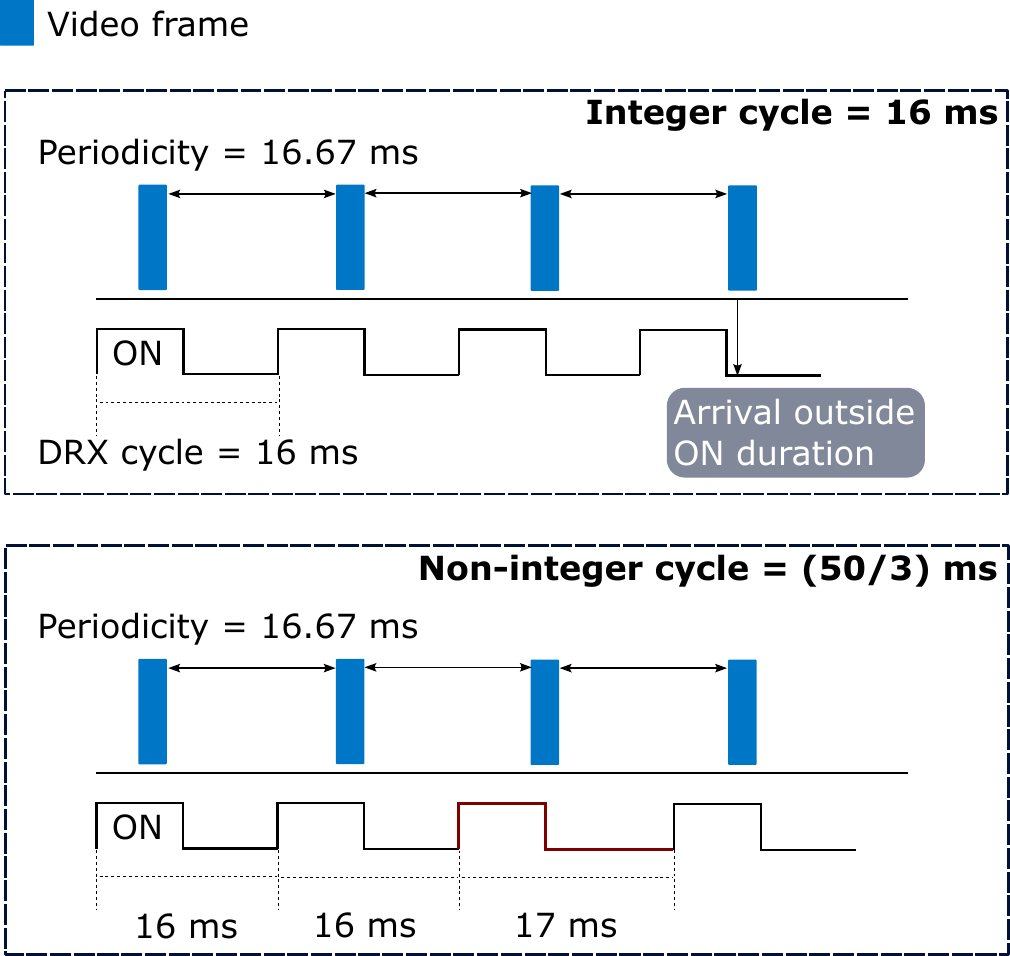}
 \caption{Illustration of non-integer cycle introduced for DRX in Release~18.}
 \label{fig:fig9}
\end{figure}

DRX operation includes also additional parameters to determine the PDCCH monitoring for DL or UL retransmission scheduling and an optional shorter DRX cycle for PDCCH monitoring that is applied for a configured time after the inactivity time has expired, before longer DRX cycle is applied for PDCCH monitoring. 

As discussed in Section~\ref{sec:traffic} and in~\cite{bl4, bl12, bl13}, the XR traffic stream packet rate depends on the assumed video frames per second (fps). This results non-integer packet inter-arrival time (i.e., 60 fps results 16.67~ms). As the in the DRX operation the PDCCH monitoring periodicity can only be configured with an integer ms granularity, the timing of the PDCCH monitoring window will not stay aligned with the XR traffic arrival. The mismatch may lead to XR frames arriving outside DRX active time and thus not being able to be transmitted further. As a result of extra delay introduced due to the mismatch, the frame might not be able to meet its PSDB requirement. To overcome this and enable more optimal DRX configuration avoiding too frequent or prolonged PDCCH monitoring, means to configure the PDCCH monitoring periodicity with rational DRX cycle lengths that match to the XR video frame rates (e.g., 60 fps) was introduced. Network may now simply configure UE with a non-integer DRX cycle from a given set of fractional DRX cycles (e.g., ms50over3 corresponding 50/3 ms as demonstrated in Fig.~\ref{fig:fig9}). 

Furthermore, as re-transmissions are not necessary for all XR related information, such as pose information, a method was introduced to disable waking up for the monitoring of potential UL re-transmissions for selected configured UL grants.

\textcolor{black}{In addition to simulation results captured in Section~\ref{sec:sim_results}, more results for various power saving techniques that were evaluated as part of Release 18 study (note, that not all were agreed to be specified) can be found in Annex B.2 TR 38.835~\cite{bl13}. Particularly, results showing benefits of aligning DRX pattern with XR traffic were captured in Annex B.2.1 TR 38.835~\cite{bl13}}.

\section{Evaluation of XR Enhancements over NR}
\label{sec:sim_results}

To demonstrate benefits of the XR enhancements introduced in Release~18, we conducted a system-level simulation campaign where some of the enhancements described in Section~\ref{sec:rel18_enhanc} were evaluated. Particularly, we use the 3GPP evaluation methodology described in~\cite{bl12, bl13} and the traffic model summarized in Section~\ref{sec:traffic}. \textcolor{black}{Note that the used stochastic XR traffic originally was derived from analysis of real XR traffic flows (see e.g.,~\cite{bl11} where SA4 conducted an extensive study and gathered information related to media and eXtended Reality traffic), and other models of the system-level simulator are also extracted from measurements (e.g., the used radio propagation models). Our system-level simulations follow the tutorial in~\cite{bl17}, where more information of the models of the simulations can be found.}

\textcolor{black}{The choice of the enhancements for this section was motivated by timescale of the enhancements, i.e., slot-level operation for BSR enhancements vs. round trip time (hundreds of ms) for L4S. We focus on the main metric of interest agreed in RAN1, which is XR capacity that depends on PDB of few milliseconds. Therefore, the following set was considered for simulation campaign:}
\begin{itemize}
\item Buffer status report enhancements.
\item	Application awareness for scheduling \textcolor{black}{with ideal DSR}.
\item	Application awareness for DRX.
\end{itemize}
\textcolor{black}{It is worth noting that all schemes were compared against baseline methods that incorporate features from Release 17 and earlier (e.g., legacy BSR and legacy DRX). Moreover, additional simulations with various parameters can be found in TR 38.838~\cite{bl12} and TR 38.835~\cite{bl13}.}

Below, we provide details of simulation framework as well as description of simulation results.

\subsection{Simulation Setup and Framework}

\begin{table}
\caption{\textbf{Simulation Parameters for FR1 InH}}
\label{table_sim}
\setlength{\tabcolsep}{3pt}
\begin{tabular}{|p{100pt}|p{115pt}|}
\hline
Parameter&
Setting \\
\hline
Channel state information (CSI)&	Periodic every 2 ms\\
Channel estimation&	Realistic with ideal CSI\\
Modulation and coding (MCS)&	Up to 256-QAM\\
Frame structure&	DDDSU\\
Downlink Scheduler&	Proportional fairness\\
Uplink Scheduler&	Adaptive transmission bandwidth\\
Target block error rate&	10\% for the first transmission\\
UE speed&	3 km/h\\
Channel model&	InH\\
Inter-site distance&	20 m\\
BS height&	3 m\\
Antenna downtilt&	90 deg\\
Carrier frequency&	4 GHz\\
Subcarrier spacing&	30 kHz\\
System bandwidth&	100 MHz\\
BS noise figure&	5 dB\\
UE noise figure&	9 dB\\
BS antenna&	32 TxRUs, 5 dBi gain\\
UE antenna&	2T/4R, 0 dBi gain\\
BS Tx power&	31 dBm\\
Max. UE Tx power ($P_{CMAX}$)&	23 dBm\\
Target RX power (P0)&	-93 dBm\\
Power scaling factor ($\alpha$)&	1 (full pathloss compensation)\\
Frame size&	Truncated Gaussian distribution,\\ 
&Mean =  Av. Data rate/ fps/ 8 [bytes]\\
&STD = 10.5\% of mean\\
&Min, max = (50\%, 150\%) of mean\\
Jitter&	Truncated Gaussian distribution,\\
&Mean = 0 ms\\
&STD = 2 ms\\
&Min, max = (-4, 4) ms\\
Frame per second&	60 fps\\
\hline
\end{tabular}
\label{tab2}
\end{table}

The scenario simulated in this section is according to~\cite{bl12, bl13} for the Indoor Hot Spot (InH) environment. Table~\ref{table_sim} shows a summary of the most relevant parameters. The UEs are dropped randomly in the environment yet ensuring that the desired number of UEs per cell are present and unchanged throughout the simulation. Simulations are run for about 10 seconds utilizing truncated Gaussian XR traffic model as described in~\cite{bl12, bl13}. The default 10 Mbps at 60 fps with 30 ms PSDB is used for UL simulations. For DL simulations the default 30/45 Mbps and 60 fps is used with 10/15 ms PSDBs. For more details on 3GPP compliant system-level simulations, we refer to a tutorial paper in~\cite{bl17}.

During Release~18 studies, the following two KPIs were considered to compare different potential solutions with each other:
\begin{itemize}
\item	XR capacity – is a maximum number of XR UEs per cell with at least 90\% of UEs being satisfied. A UE is assumed to be satisfied if more than 99\% of frames/application layer packets are successfully transmitted within its PSDB.
\item	UE power saving gain – is a ratio between the average UE power consumption with power saving technique and the average UE power consumption when UE is always on meaning no power saving technique is enabled and UE is always monitoring control channels.
\end{itemize}

In the next sub-section, we model these two KPIs among others to show benefits of enhancements in terms of capacity and power saving.

\subsection{Simulation Results}

\subsubsection{BSR Enhancements}

\begin{figure}[!b]
 \centering
 \includegraphics[width=0.95\columnwidth]{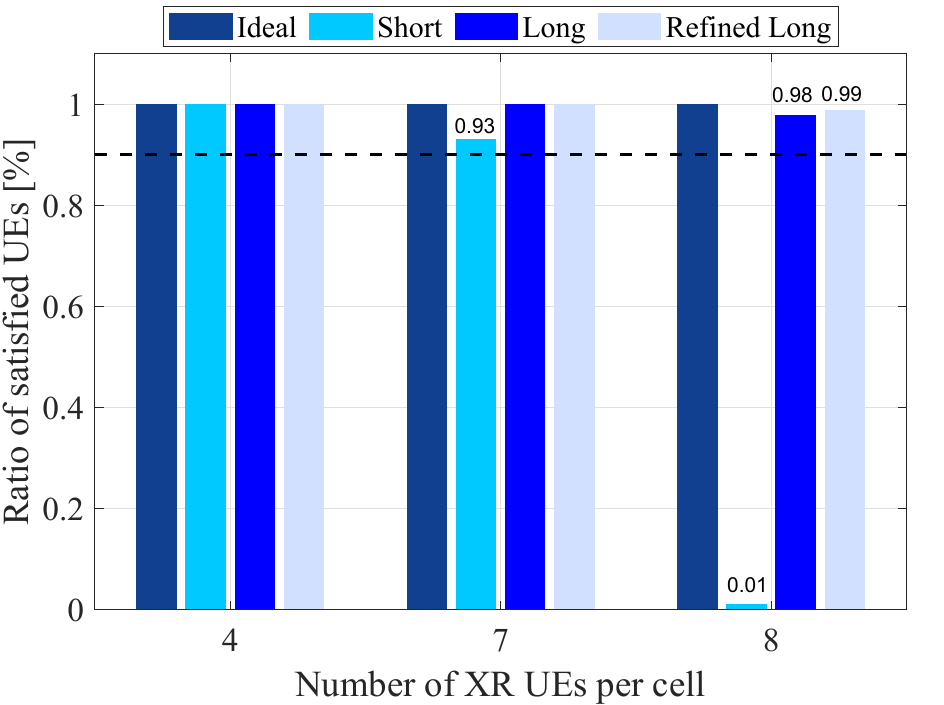}
 \caption{Ratio of satisfied XR UEs for different BSR formats in UL FR1 InH, 10 Mbps and 30 ms PSDB.}
 \label{fig:fig10}
\end{figure}

We start with evaluation of the BSR enhancements described in Section~\ref{sec:capacity_enh}. Fig.~\ref{fig:fig10} shows the impact of different BSR tables on a ratio of satisfied XR UEs in a 3GPP InH environment. This environment is characterized by high signal-to-interference-plus-noise ratio (SINR) conditions and the primary XR capacity limiting factor is the cell capacity. The 3GPP Short, Long and Refined Long BSR types are simulated with a reference to an ideal case, where the gNB has instant and perfect knowledge of the UE’s BS status (Ideal BSR in Fig.~\ref{fig:fig10}). The main difference between legacy Short and Long BSR tables is in the quantization of the buffer status values, where for Short BSR table, due to limited number of available BS values, the gap between each BS value is much larger than in Long BSR table.

From Fig.~\ref{fig:fig10}, we observe that as the number of XR UEs per the cell increases it reaches a limit of 8 XR UEs per cell. Given the high data rate of XR services this results in median resource block (RB) utilization of over 99.5\% for the cells. Introducing BSR overhead, leads to a total degradation in terms of ratio of satisfied UEs for the least efficient Short BSR. The cells then drop to 7 satisfied XR UEs. Long and Refined BSR do introduce some reduction of ratio of satisfied XR UEs, but the cells can still support 8 UEs. The primary reason for these effects is the overhead introduced by the realized BSR quantization error in an already loaded cell.

\begin{figure}[!t]
 \centering
 \includegraphics[width=0.90\columnwidth]{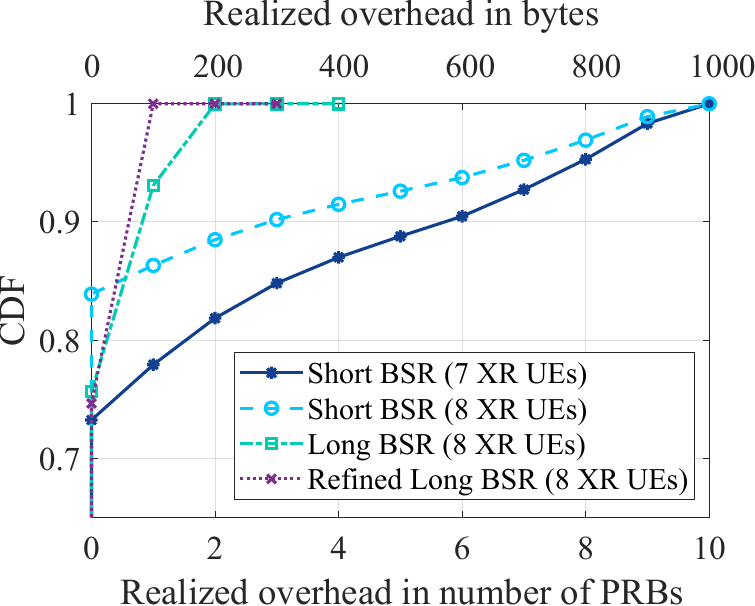}
 \caption{CDF of realized overhead in terms of PRBs and bytes from different BSR formats in UL FR1 InH, 10 Mbps and 30 ms PSDB.}
 \label{fig:fig11}
\end{figure}

Fig.~\ref{fig:fig11} shows the realized overhead for the various BSR types. The realized overhead is the number of padding bytes added to the transport block to match the overallocated resources by the gNB, due to upper bound reporting by the UE. The cumulative distribution function (CDF) starts from around 70\% because the statistics collected include transport block transmissions with no padding. This means that most of the time BSR reporting overhead is not in fact realized overhead and only last transmissions, towards emptying the LCG buffer, lead to realized BSR overhead. It also shows the effect of buffer build up for the 8 XR UEs case with Short BSR, where this leads to XR happiness collapse. The conversion from bytes of overhead to number of PRBs is approximate given the current allocation’s modulation and coding scheme (MCS).

\begin{figure}[!b]
 \centering
 \includegraphics[width=0.95\columnwidth]{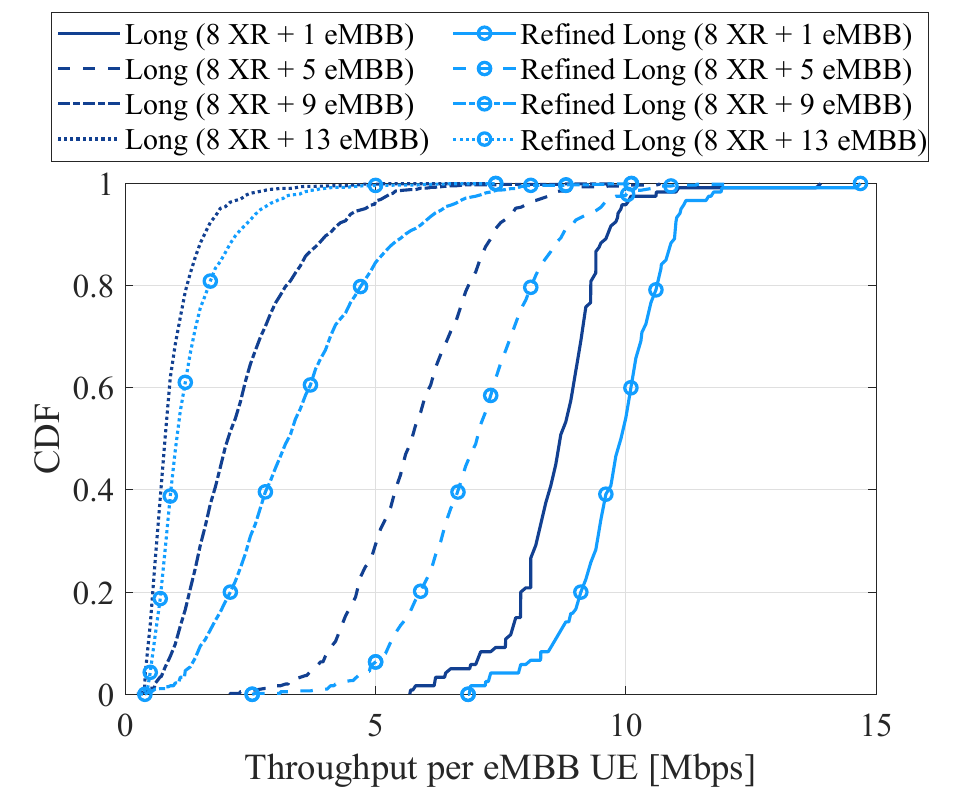}
 \caption{CDF of FTP3 throughput for Refined Long BSR and legacy Long BSR for eMBB users in a mixed XR + eMBB users scenario.}
 \label{fig:fig12}
\end{figure}

We further note, that in both Fig.~\ref{fig:fig10} and Fig.~\ref{fig:fig11} only XR users were deployed in the network, and we could add extra satisfied XR users if the cell had spare capacity. 

Next, we evaluate a scenario where in addition to XR users, some enhanced mobile broadband (eMBB) users are also present. eMBB users are modelled as having FTP3 traffic \textcolor{black}{(the largely used in 3GPP model for generic eMBB type of simulations)~\cite{bl17, 3GPP_36889, 3GPP_36814}} with 1 Mbps offered load, generating 0.125 Mbytes files every second. Fig.~\ref{fig:fig12} shows the CDF of throughput per eMBB UE in the case of mixed traffic. \textcolor{black}{In this simulation, XR users are given hard priority and scheduled first, while eMBB users are served as best effort within the remaining resources every TTI.} In a cell where most resources are already allocated to XR users, the impact of realized overhead of the prioritized XR is clearly visible as FTP3 traffic is significantly delayed. The comparison in Fig.~\ref{fig:fig12} is between the legacy Long BSR and the, newly Release~18 defined, Refined Long BSR table. Fig.~\ref{fig:fig11} and Fig.~\ref{fig:fig12} demonstrate that the lower realized overhead from Refined Long BSR helps to improve the throughput of eMBB users as compared to Long BSR table.



\subsubsection{Application Aware Scheduling Using PDU Set Information}

From scheduler’s point of view, information about PDU sets can be exploited to better prioritize the XR users for their upcoming DL/UL transmissions. Most of the legacy schedulers~\cite{bl18} do not use packet level information to evaluate their metrics and mostly rely on long term and flow-level statistics, e.g., average throughput, to decide which user gets to transmit/receive data next. We propose to use the information regarding the buffering time of a PDU set payload instead of focusing on individual PDU delays. Moreover, it is suggested to schedule the PDUs of a PDU set consecutively, instead of sending several PDUs from different PDU sets. These considerations are essential to satisfy the PDU set level performance requirements and thus, achieving a better user experience.  

In the following, we study a use case for a PDU set aware DL scheduler which uses two key information elements of PDU set size and PDU set arrival time. These information elements are then translated into remaining PDU set delay budget (PSDB) and remaining buffered PDU set data. Moreover, it is aware of PDU set sequence numbers, which is used by the scheduler to identify PDUs of a certain PDU set. 

At each transmission time interval (TTI), the scheduler first prioritizes XR users over other traffic-type of users. Next, among the XR users (if there are multiple queued XR users), it prioritizes the ones with shorter remaining PSDB and smaller buffered data. \textcolor{black}{This is done by means of a new scheduling metric which grows exponentially with the ratio of transmitted bits to the size of the PDU set. The metric also has a linear-inverse relation to the remaining PSDB. The discarding of packets is disabled for this scenario. For the packets buffered beyond the PSDB, the scheduling priority is set to a predefined minimum value and they still get a chance to be scheduled.}

\begin{figure}[!t]
 \centering
 \includegraphics[width=0.95\columnwidth]{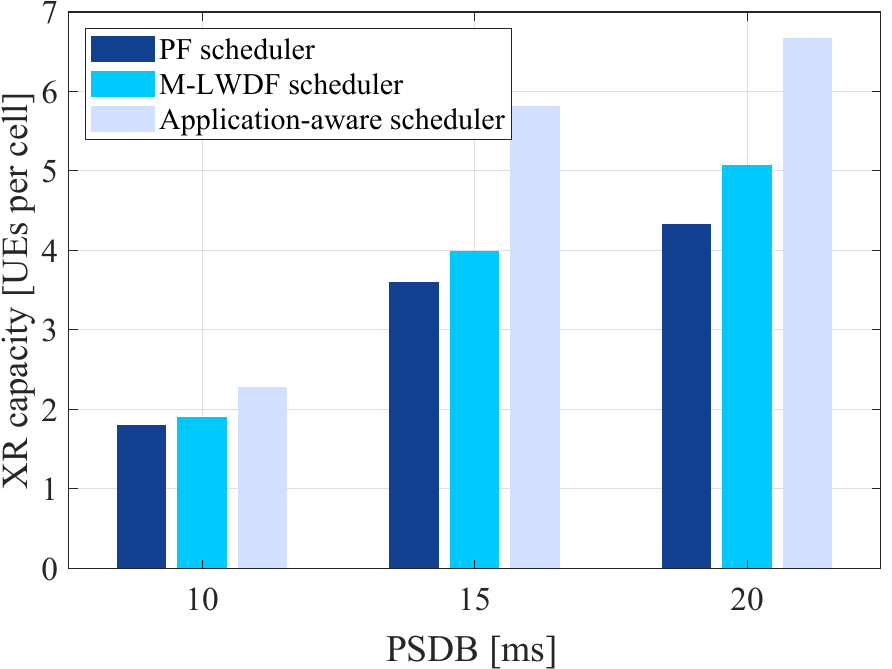}
 \caption{XR capacity with different scheduling policies in DL FR1 InH, 45 Mbps: Proportional fair (PF), modified-largest weight delay first (M-LWDF), scheduling with application awareness.}
 \label{fig:fig13}
\end{figure}

For the simulations, the eMBB users are modeled with simple full-buffer traffic to create a practical scenario. \textcolor{black}{For the delay awareness (for both application-aware and M-LWDF) the information is obtained via an ideal DSR where reporting delays and quantization errors are ignored.} More details on the logic applied at the scheduler to implement PDU set awareness are described in~\cite{bl19}. Fig.~\ref{fig:fig13} shows the XR capacity (number of satisfied XR users) versus the PSDB for different schedulers. As can be observed, the new scheduler has a better performance compared to the well-known proportional fair (PF) and the modified-largest weight delay first (M-LWDF) schedulers. The main reason is the ability of the scheduler to recognize PDUs of a set and adjust its metric based on the remaining PSBD and PDU set size that are key factors in XR user satisfaction.

\subsubsection{Power Saving Enhancements} 

Next, we discuss the performance evaluation of the enhancements that aim to improve power saving, another important metric for XR devices. As discussed in Section~\ref{sec:ps_enh}, DRX is one of the power saving mechanisms allowing UE to go to sleep mode and reduce its power consumption. 

In Release 18 MAC specifications, DRX enhancements focus on two aspects: (i) introducing non-integer DRX cycle to match the rational periodicity of XR traffic due to the typical frame rate of video codecs, and (ii) improving the DRX formula to avoid sudden drift between the DRX cycle and XR traffic when the SFN wraps around. 

\begin{figure}[!t]
\centering
    \subfloat
    {
        \includegraphics[width=0.43\textwidth]{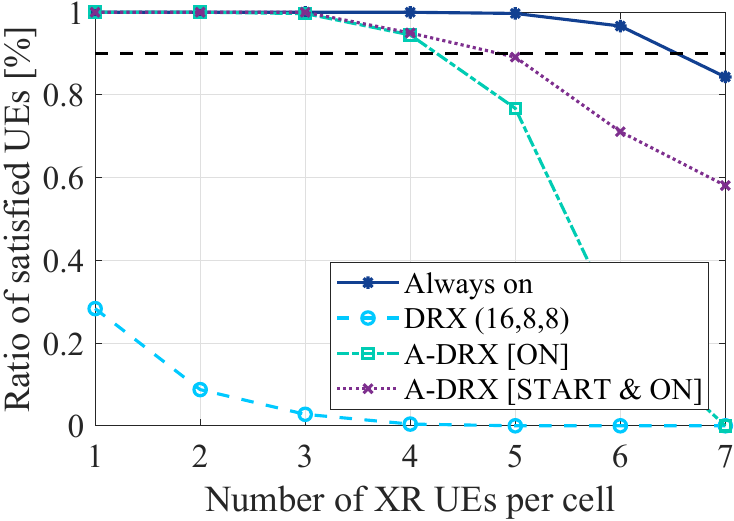}
        \label{fig:fig14a}
    }
		\vspace{5mm}
    \quad
		
    \subfloat
    {
        \includegraphics[width=0.43\textwidth]{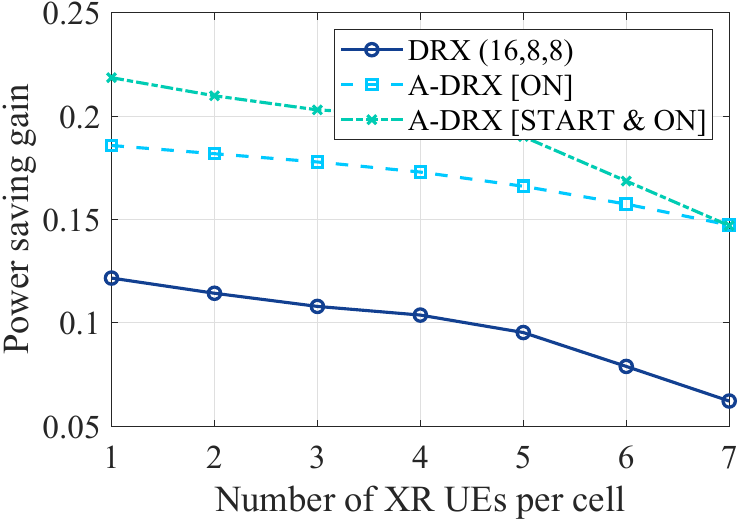}
        \label{fig:fig14b}
    }
    \quad
    \caption{Comparison of power saving schemes in DL FR1 InH 30 Mbps, 10 ms PSDB in terms of satisfied UEs and power saving gain: (i) legacy DRX (16, 8, 8) vs (ii) DRX with application awareness.
    \label{fig:fig14}
}
\end{figure}

The need of introducing non-integer periodicity in DRX comes from the evidence that a small mismatch between the XR traffic periodicity and DRX cycle causes the XR traffic to drift away from the UE active period. Even with a DRX integer configuration closely matching the XR traffic periodicity, including an ON duration and an inactivity timer long enough to cover a full DRX cycle to account for jitter, the XR traffic performance are highly affected. For example, in Fig.~\ref{fig:fig14} the curve labeled DRX (16,8,8) corresponds to a DRX configuration with: (i) cycle of 16 ms, (ii) ON duration of 8 ms, and (iii) inactivity timer of 8 ms. The curve labeled ``Always ON'' represents instead the case when power saving features are disabled and UEs are always monitoring the channel. We observe that XR traffic is generated at 60 fps, resulting in a mismatch of only 0.167ms. This mismatch accumulates over time causing the XR traffic to fall outside the ON duration. When this happens, the traffic is delayed until the next DRX cycle starts, since the UE is supposed to be inactive if no traffic has been scheduled during the On duration. As illustrated in the figure, the ratio of satisfied XR UEs quickly drops from less than 40\% to 0\% when the cell load is increased.

\begin{figure*}[!t]
 \centering
 \includegraphics[width=0.95\textwidth]{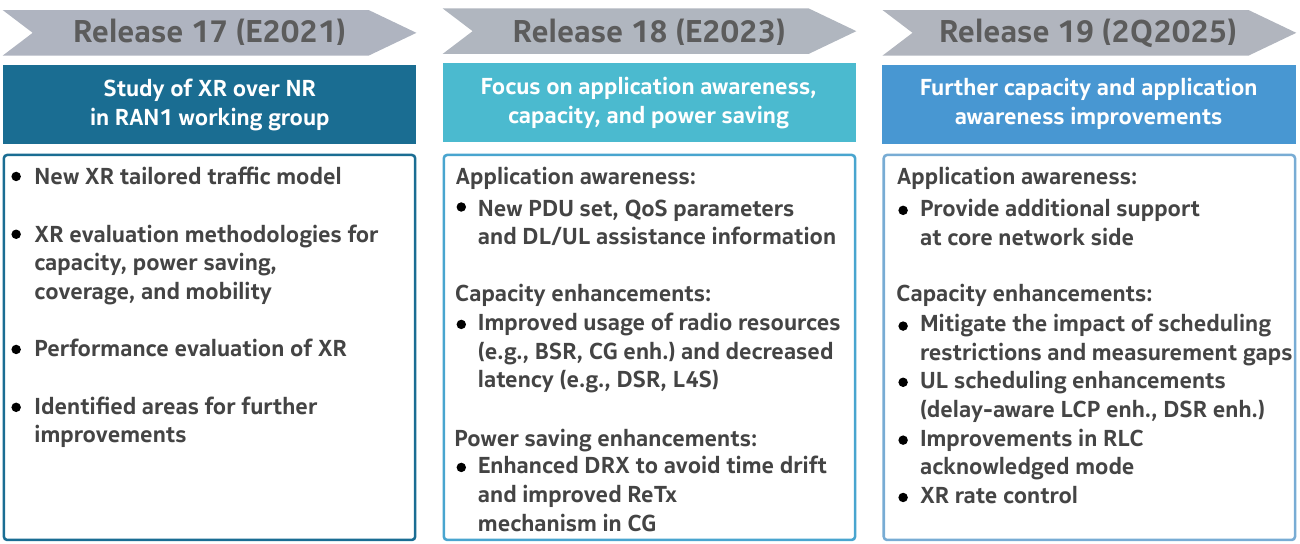}
 \caption{\textcolor{black}{Summary of 3GPP progress on XR over Release 17, 18, and 19.}}
 \label{fig:fig15}
\end{figure*}

While Release 18 DRX enhancements solve the mismatch issue, the configuration of the starting slot and ON duration of the DRX cycle needs to be optimized. In particular, a starting slot that is not synchronized with the expected XR frame arrivals would require an ON duration longer enough to cover this gap and the jitter, thus reducing the UE power saving. In contrast, when the starting slot of the DRX cycle is configured to match the expected frame arrival, the ON duration can be dynamically adjusted to minimize the UE power consumption while fulfilling the QoS requirements. \textcolor{black}{To this end, we model the problem of minimizing the UE power consumption over time subject to UE satisfaction as a stochastic optimization problem. To solve this problem, we derive Adaptive DRX (A-DRX) using the drift-plus-penalty method, where we defined the Lyapunov drift and penalty to model the drifting of the UE satisfaction and the UE power consumption, respectively~\cite{ADRX_Stefano}. A-DRX dynamically learns the best parameters of the DRX configuration using the information related to the PDU set and End of Data Burst specified in Release 18, which helps to evaluate the drifting of the UE satisfaction due to XR packets delivered beyond the PSDB and the penalty paid when keeping the UE awake more than needed to deliver a XR packet.} The addition of the adaptive policy on top of the Release 18 DRX enhancements increases the performance both in terms of XR capacity and power saving gain. As illustrated in Fig.~\ref{fig:fig14}, A-DRX improves the ratio of satisfied XR UEs closely approaching the case when UEs are always active while still achieving a significant power saving gain. In particular, when the policy can only adapt the ON duration (label A-DRX [ON]), at most 4 UEs are served and the power saving gain range between 15\% and 18\%, while legacy DRX barely achieves 10\% power saving gain without being able to fulfill the QoS requirements of any XR user. Instead, the possibility of adapting the starting slot in addition to the ON duration results in an additional satisfied UE and a power saving gain up to 23\%, which is close to the maximum bound of 25\% that can be obtained with a jitter that spans half the traffic periodicity.

\section{Outlook of XR Enhancements in Release 19}
\label{sec:rel19_enhanc}

The work towards even better end-to-end XR user experience and higher system capacity continues in 5G-Advanced Release 19 as outlined in the recent white paper here~\cite{bl20}. In addition, the detailed objectives of the RAN and SA Release 19 items on the next phase XR enhancements appear in~\cite{bl21} and~\cite{bl22}, respectively. \textcolor{black}{Release 19 XR enhancements mostly focus on improving capacity of XR in DL and UL directions.}

\textcolor{black}{Among others, scheduling restrictions and measurement gaps arising from UEs performing RRM measurements were found to reduce the XR capacity significantly as reported in~\cite{bl13}. In order to address such challenges, in Release 19 new  mechanism for selective skipping of RRM measurements occasions is defined that would otherwise prevent the gNB from scheduling latency critical XR payloads. The studies related to intra-UE multi-modality to facilitate efficient support for XR applications with multiple QoS flows was conducted. In addition, related UL scheduling enhancements to further increase XR capacity are also in the Release 19 scope, i.e., delay-aware logical channel priority enhancements to resolve the issue of data with low remaining time being delayed due to data from other LCHs with no delay critical data.} 

\textcolor{black}{Another enhancement is aiming for RLC acknowledged mode to make it feasible for XR case with limited PSDB. Enabling the use of RLC acknowledged mode for XR services with PSDB’s as low as 10 ms would help improve the reliability for cases where the lower layer HARQ fails (e.g., due to errors on physical layer ACK/NACK feedback).} 

\textcolor{black}{To deal with congestion in UL, Release 19 is also looking at XR rate control mechanisms. The goal is to provide MAC layer XR rate control signaling to enable faster source rate adaption to uplink congestion.}

Furthermore, the framework for XR application-awareness based on the introduction of PDU set QoS attributes, will be further refined in Release 19 \textcolor{black}{for the core network side}. This includes potential options for new 5G QoS Identifiers (5QIs) to improve PDU set and Data Burst based QoS handling as well as related signaling and procedures between the application, the core network and the RAN. One special area of interest has been to improve how the applications can provide to the network more information about their PDU sets, Data Bursts or updated traffic characteristics in a more accurate and secure manner than possible in Release 18. 

As outlined in~\cite{bl20}, there are many other items in 5G-Advanced Release 19 scope that will contribute to improved XR performance. To mention a few, the MIMO evolution will continue to improve the spectral efficiency (and thereby also the XR system capacity) by enabling support for \textcolor{black}{larger number of} transmit ports by introducing enhancements to the channel state information (CSI) framework. Innovations related to conditional mobility enhancements with shorter interruption times and lower layer triggered mobility (LTM) for dual connectivity will be considered, offering benefits for latency critical XR services that dislike interruptions during handovers. Finally, Release 19 also includes a study on application enablement for Localized Mobile Metaverse Services~\cite{bl23} as will be conducted in SA working group six.

\begin{table}
\caption{\textbf{List of Abbreviations}}
\label{tab1}
\setlength{\tabcolsep}{3pt}
\begin{tabular}{|p{60pt}|p{155pt}|}
\hline
Abbreviation&
Description \\
\hline
3GPP&	3rd generation partnership project\\
5G&Fifth Generation\\
5GC&	5G core\\
5GS&	5G system\\
AF&	Application function\\
AMF&	Access and mobility management function\\
AR&	Augmented reality\\
AUSF& Authentication Server Function\\ 
BSR&	Buffer status report\\
CDF&	Cumulative distribution function\\
CG&	Configured grant\\
DL&	Downlink\\
DRB&	Data radio barrier\\
DRX&	Discontinuous Reception\\
DSR&	Delay status report\\
ECN&	Explicit Congestion Notification\\
eMBB&	Enhanced Mobile broadband\\
EoDB&	End of data burst\\
fps&	Frame per second\\
gNB&	gNodeB\\
GTP-U&	GPRS Tunnelling Protocol for the user plane\\
IETF&	Internet Engineering Task Force\\
IP&	Internet protocol\\
L4S&	Low Latency Low Loss Scalable Throughput\\
LCG&	Logical channel group\\
MAC&	Medium access control\\
MR&	Mixed reality\\
NAS& Non-Access Stratum\\
NEF&	Network exposure function\\
NRF& NR Repository Function\\
NSSF& Network Slice Selection Function\\
PCF&	Policy control function\\
PDB&	Packet delay budget\\
PDCCH&	Physical Downlink Control Channel\\
PDCP&	Packet data convergence channel\\
PDU&	Protocol data unit\\
PER&	Packet error rate\\
PHY&	Physical layer\\
PRB&	Physical resource block\\
PSDB&	PDU set delay budget\\
PSER&	PDU set error rate\\
PSI&	PDU set importance\\
PSIHI&	PDU Set Integrated Handling Indicator\\
PUSCH&	Physical Uplink Shared Channel\\
QoS&	Quality of Service\\
RAN&	Radio access network\\
RB&	Resource block\\
RLC&	Radio link control\\
RRC&	Radio resource control\\
RTCP&	RTP control protocol\\
RTP&	Real Time Transport Protocol\\
SA&	Service and System Aspect\\
SDAP& Service Data Application Protocol\\
SMF&	Session management function\\
TCP&	Transmission Control Protocol\\
TSCAI& Time Sensitive Communication Assistance Information\\
UAI&	UE assistance information\\
UDM& Unified Data Management\\
UDR& Unified Data Repository\\
UE&	User equipment\\
UL&	Uplink\\
UPF&	User plane function\\
UTO-UCI&	Unused transmission occasion – uplink control information\\
VR&	Virtual reality\\
XR& Extended reality\\
\hline
\end{tabular}
\label{tab1}
\end{table}

\section{Conclusion}
\label{sec:conclusion}

3GPP continues driving the development of 5G-A to accommodate the requirements introduced by XR services. The work is spread among many working groups inside 3GPP to provide seamless end-to-end connectivity to support various XR applications. In this paper, we focused on physical and higher layer enhancements introduced for NR (summarized in Fig.~\ref{fig:fig15}) that are taking into account the specifics of XR traffic. Furthermore, the overview provides an insight on system architecture development that impacted radio access part. 

In addition to the overview of the enhancements introduced in Release 18, we conducted a system-level simulation campaign to evaluate few selected enhancements. Particularly, we demonstrated the benefits of newly introduced Refined Long BSR and application awareness used in scheduling policy in terms of capacity. It was shown that Refined Long BSR provides less overhead as compared to the earlier BSR table formats resulting in higher capacity for XR users and eMBB users in case of mixed traffic scenario. Additionally, application awareness for scheduling improved the capacity by exploiting the knowledge about PDU sets and its metrics such as PSDB. Moreover, the application awareness was also considered beneficial to increase power saving by improving decision on a proper selection of parameters for DRX.

Finally, we highlighted the main aspects of ongoing items related to XR in Release 19 since the work to further improve XR service provided by 5G-A continues in 3GPP. \textcolor{black}{Summarizing, RAN working groups continue developing features to further improve XR capacity. In addition, Release 19 is aiming to provide support of application awareness from core network. Moreover, besides features developed under XR umbrella there are various other technological components that may help to provide a better XR service such as MIMO evolution and lower layer triggered mobility to name a few. In addition to those, areas like Artificial Intelligence (AI) and Machine Learning (ML), enhanced XR positioning and many others areas could be further considered to improve XR performance.}

\section*{Acknowledgment}
The authors would like to thank Claudio Rosa, Santiago Morejon, Jian Song, Zexian Li, Pouria Paymard, Tero Henttonen, Devaki Chandramouli, Antti Toskala and other Nokia colleagues for their enormous support and contribution.


\bibliographystyle{ieeetr}
\bibliography{final_version}
\vspace{-10mm}

\begin{IEEEbiography}[{\includegraphics[width=1in,height=1.25in,clip,keepaspectratio]{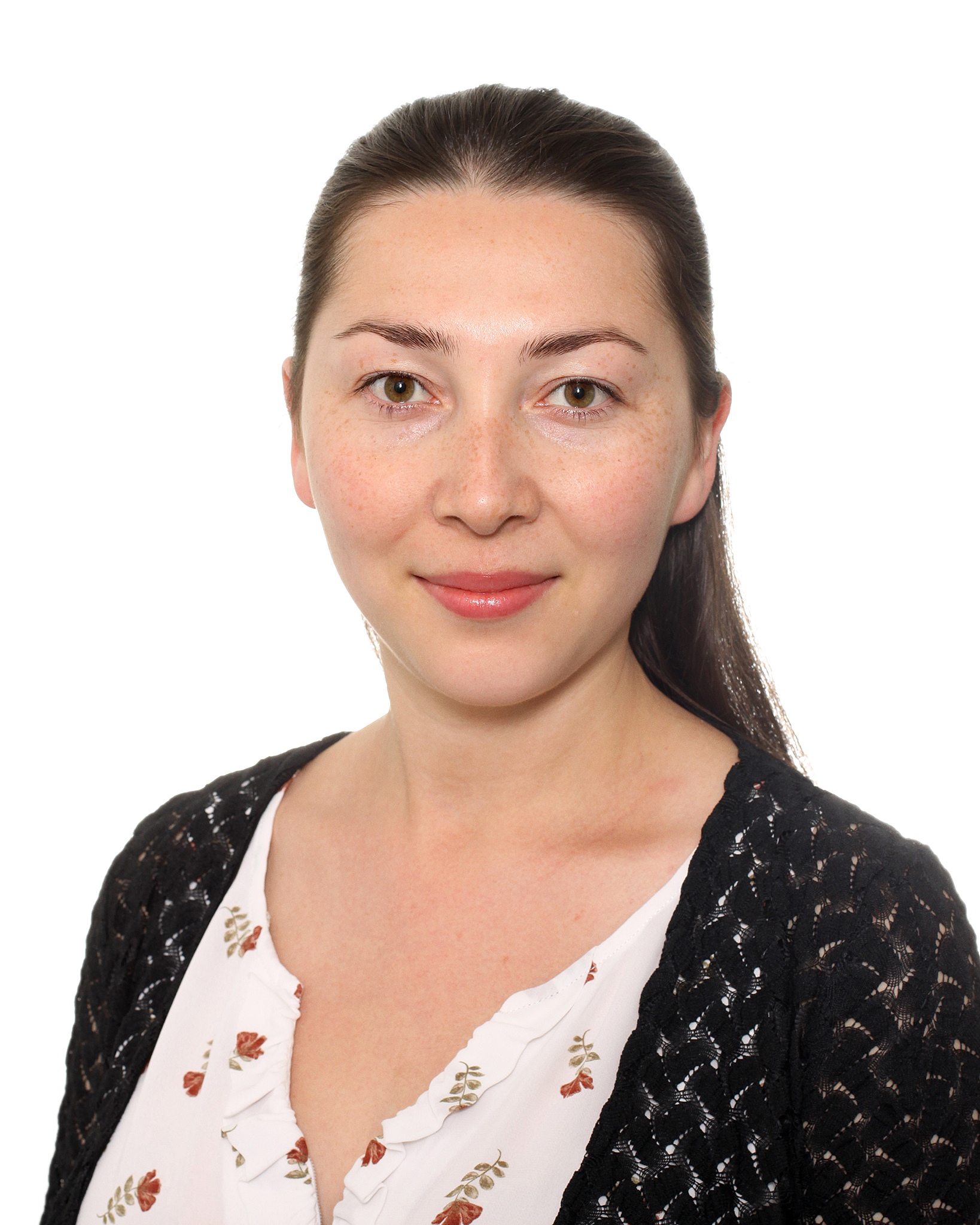}}]{Margarita Gapeyenko} is a Senior Standardization Specialist and a 3GPP RAN1 Delegate at Nokia Standards. She received her M.Sc. from the University of Vaasa, Finland, in 2014 and her Ph.D. from Tampere University, Finland, in 2022. Her research interests include millimeter wave networks, UAV communications, XR, 5G-Advanced, and 6G wireless systems.  
\end{IEEEbiography}
\vspace{-7mm}
\begin{IEEEbiography}[{\includegraphics[width=1in,height=1.25in,clip,keepaspectratio]{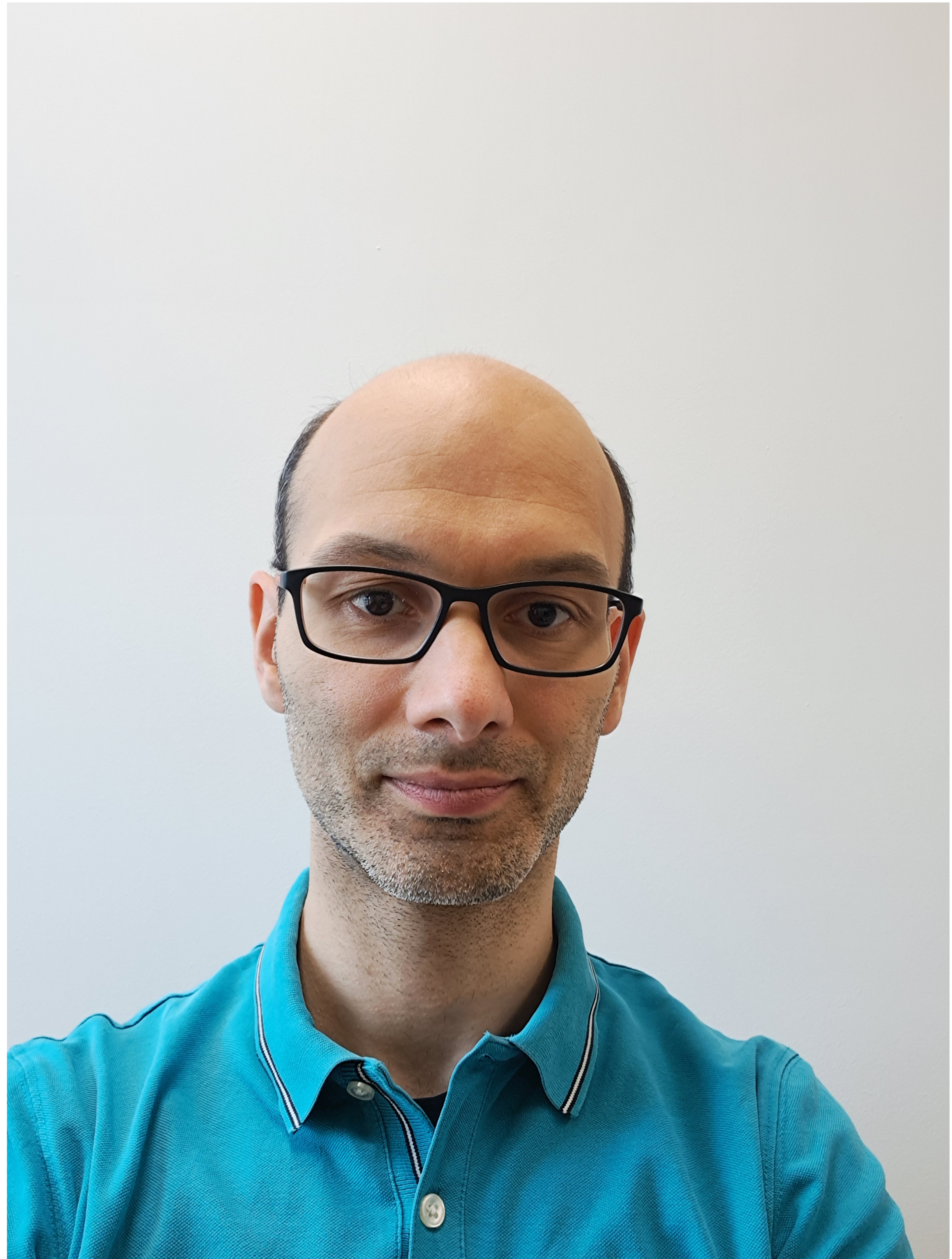}}]{Stefano Paris} is a Senior Research Scientist at Nokia Standards. He received his B.Sc. (2004) and M.Sc. (2007) from University of Bergamo, and his Ph.D. (2011) from Politecnico di Milano. His research interests include resource allocation, optimization, artificial intelligence and machine learning applied to wireless and wired networks.
\end{IEEEbiography}
\vspace{-7mm}
\begin{IEEEbiography}[{\includegraphics[width=1in,height=1.25in,clip,keepaspectratio]{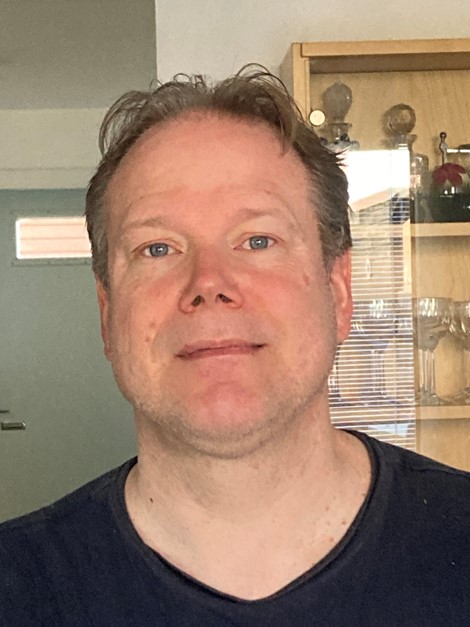}}]{Markus Isomäki} is a Principal Research Lead at Nokia Standards. He received is M.Sc. in 1998 from Helsinki University of Technology. One of his research interests is supporting delay sensitive applications in 5G advanced and 6G networks.
\end{IEEEbiography}
\vspace{-7mm}
\begin{IEEEbiography}[{\includegraphics[width=1in,height=1.25in,clip,keepaspectratio]{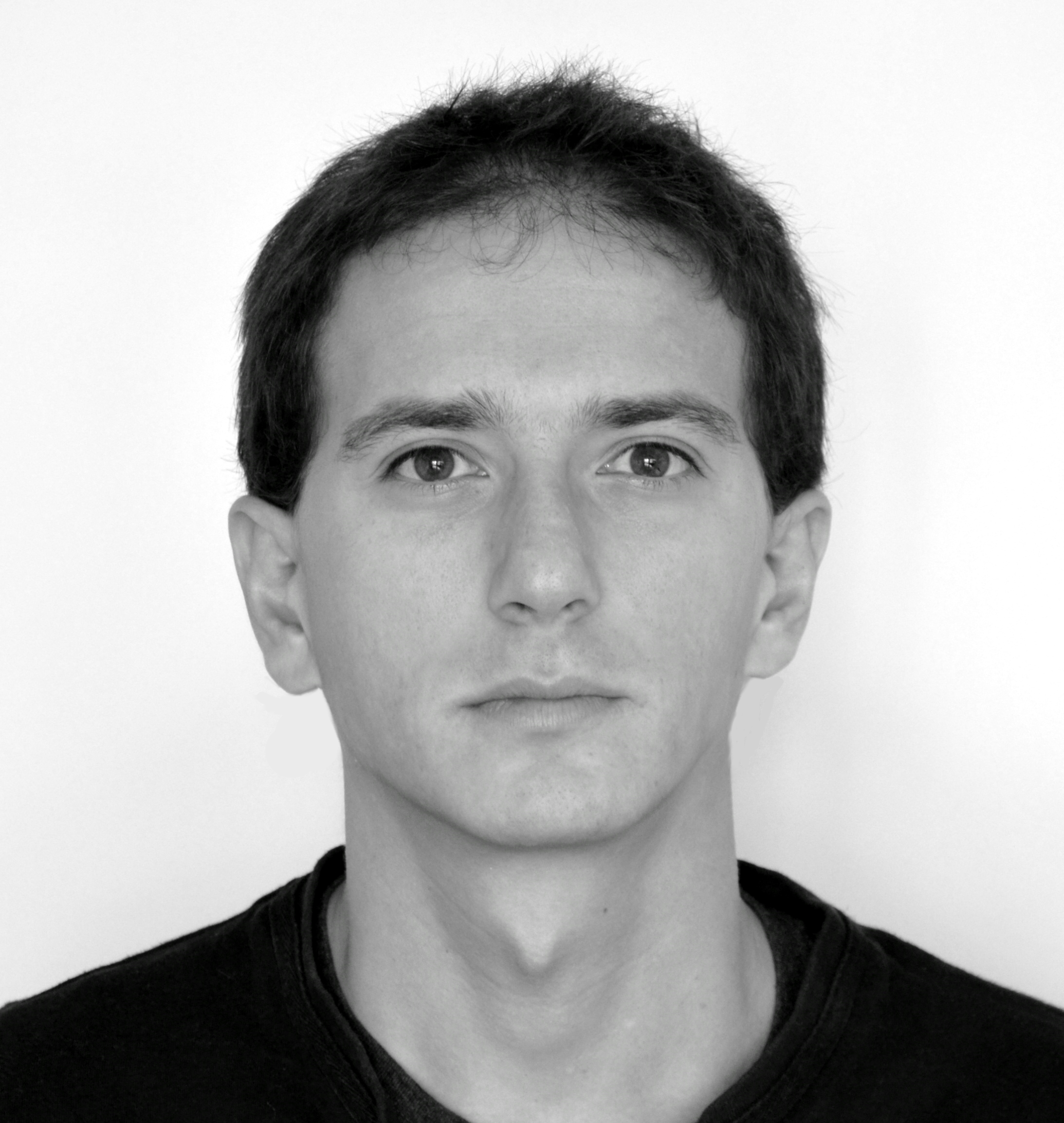}}]{Boyan Yanakiev} is a Senior standardization specialist at Nokia Standards. He received his B.Sc. (2006) from Sofia University and M.Sc. (2008) and Ph.D. (2011) from Aalborg University. He is currently focusing on 5G advanced and 6G topics for XR optimizations in RAN.
\end{IEEEbiography}
\vspace{-7mm}
\begin{IEEEbiography}[{\includegraphics[width=1in,height=1.25in,clip,keepaspectratio]{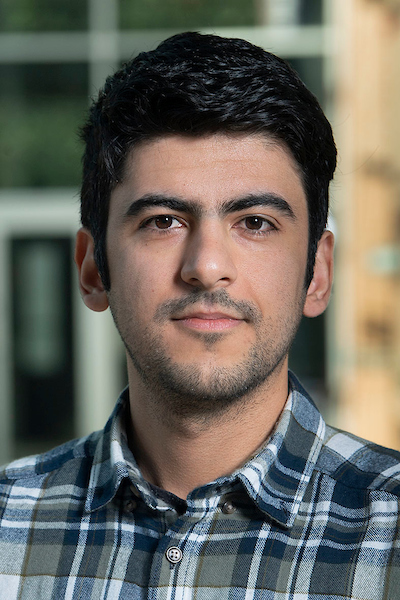}}]{Abolfazl Amiri} is a Senior standardization specialist at Nokia Standards. He received his Ph.D. in telecommunications from Aalborg University in 2022. His main research interests include 5G/6G standards, radio access protocols, and XR. 
\end{IEEEbiography}
\vspace{-7mm}
\begin{IEEEbiography}[{\includegraphics[width=1in,height=1.25in,clip,keepaspectratio]{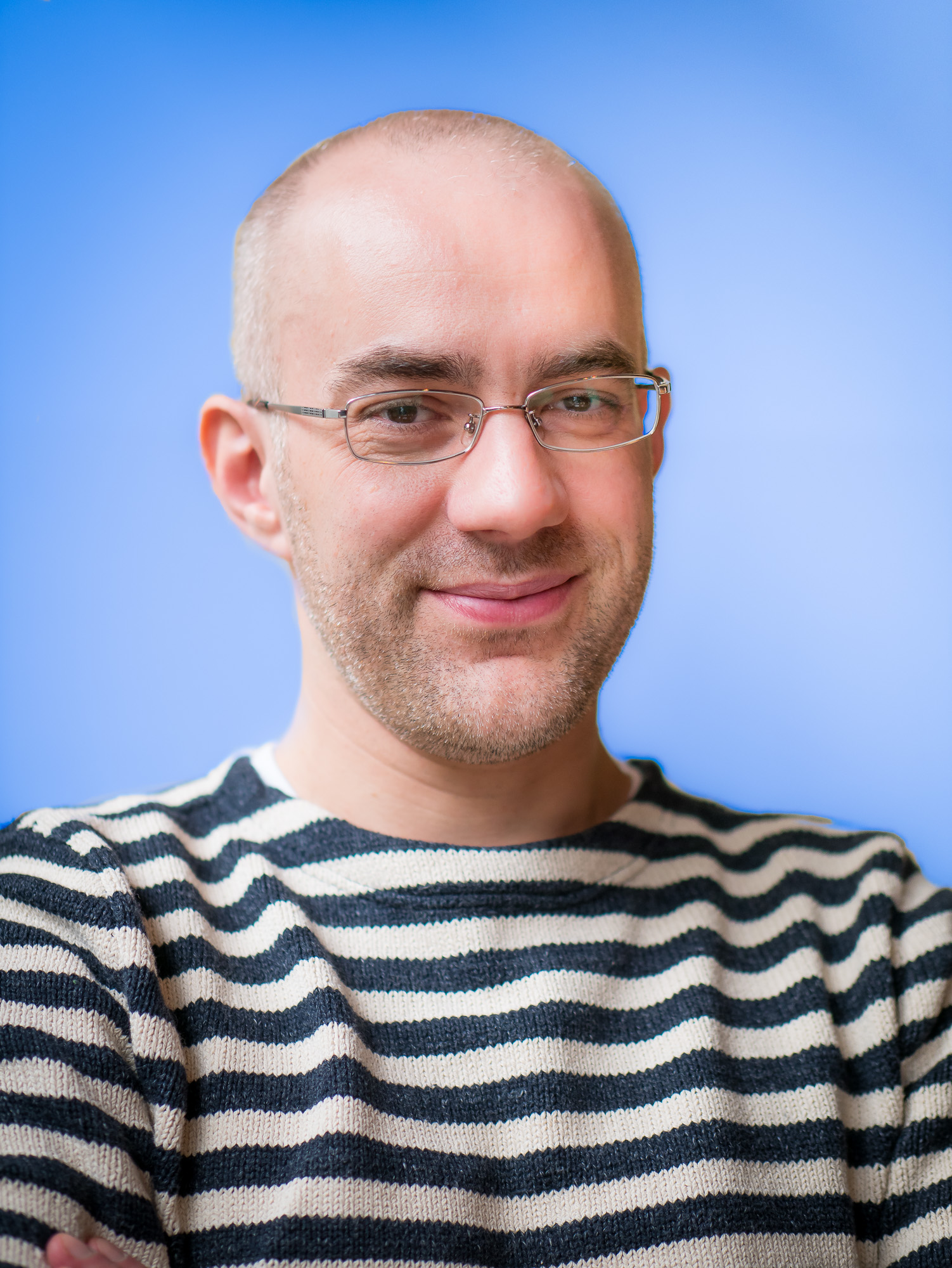}}]{Benoist Sébire} is a Principal Standardization Lead at Nokia Standards. He received his M.Sc. in Electrical, Electronics and Communications Engineering from the National School of Applied Science and Technology in France, in 1997. He has been contributing to standardization work in 3GPP since 2000 and his research interests include radio protocols, XR and is currently investigating possible enhancements for 6G wireless systems. 
\end{IEEEbiography}
\vspace{-7mm}
\begin{IEEEbiographynophoto}{Jorma Kaikkonen} received the M.S.E.E degree from the University of Oulu, Oulu, Finland in 1999. He joined Nokia in 1996. Currently he is working at Nokia Standards, Oulu, Finland, where he is working on various topics related to radio research and standardization. 
\end{IEEEbiographynophoto}
\vspace{-7mm}
\begin{IEEEbiography}[{\includegraphics[width=1in,height=1.25in,clip,keepaspectratio]{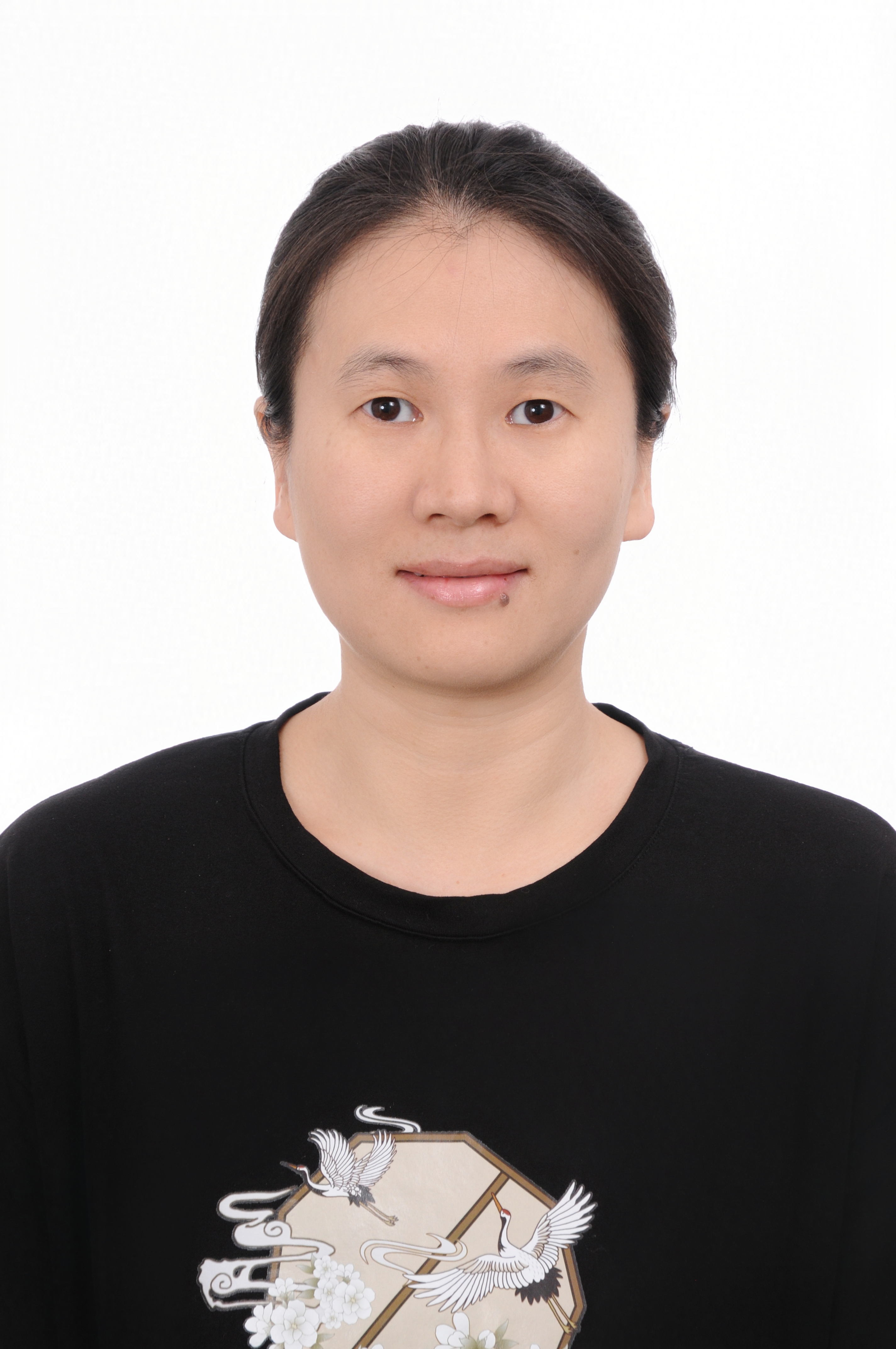}}]{Chunli Wu} is a Principal Standardization Lead at Nokia Standards. She received her M.Sc. in Communication Engineering from Beijing University of Post and Telecommunication in China, in 2007. She has been contributing to standardization work in 3GPP since her graduation and her research interests include radio protocols for 4G/5G/6G, XR, network energy saving and UE power saving, etc. 
\end{IEEEbiography}
\vspace{-7mm}
\begin{IEEEbiography}[{\includegraphics[width=1in,height=1.25in,clip,keepaspectratio]{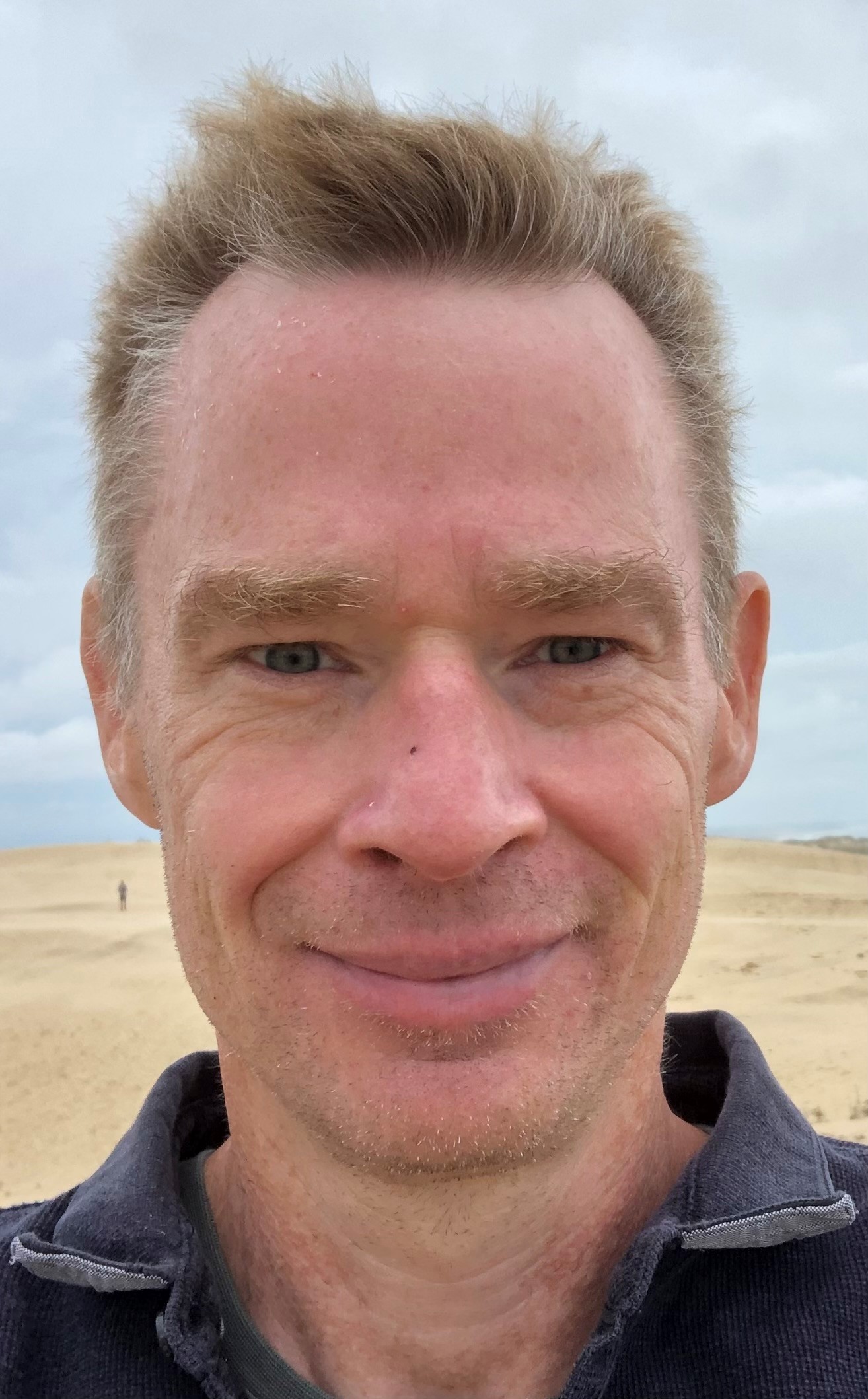}}]{Klaus I. Pedersen} is a Nokia Bell Labs Fellow, leading the Radio Access Research Team in Aalborg, and an external professor at Aalborg University. He received his M.Sc. degree (1996) and his Ph.D. (2000) from Aalborg University, Denmark. His current research covers access protocols and radio resource management for 5G-Advanced and 6G systems.
\end{IEEEbiography}
\vspace{-7mm}

\balance

\EOD

\end{document}